\documentclass[preprint,12pt,authoryear]{elsarticle}

\usepackage{mathrsfs}
\usepackage{amsmath}
\usepackage{amssymb}
\usepackage{dsfont}
\usepackage{xcolor}
\usepackage[normalem]{ulem}
\usepackage{url}
\def\SX{{\mathscr X}}
\def\Ex{{\rm I}\!{\rm E}}
\def\Var{\mbox{\rm Var}}
\def\Cov{\mbox{$\mathds{C}$\rm ov}}
\def\N{{\rm I}\!{\rm N}}

\newcommand{\W}[1]{{\color{blue} #1}}

\journal{Spatial Statistics}
\newpageafter{abstract}

\begin{document}

\begin{frontmatter}

\title{A criterion and incremental design construction for simultaneous kriging predictions}
\author[1]{Helmut Waldl\corref{cor1}}
\ead{helmut.waldl@jku.at}
\author[1]{Werner G. Müller}
\ead{Werner.Mueller@jku.at}
\W{\author[2]{Paula Camelia Trandafir}
\ead{camelia@unavarra.es}}
\cortext[cor1]{Corresponding author}
\affiliation[1]{organization={Department of Applied Statistics, Johannes-Kepler-University Linz},
            addressline={Altenberger Stra{\ss}e 69},
            city={Linz},
            postcode={A-4040},
            country={Austria}}

\affiliation[2]{organization={Department of Statistics, Informatics and Mathematics, Public University of Navarre},
            addressline={Campus Arrosadia},
            city={Pamplona},
            postcode={31006},
            country = {Spain}}

\begin{abstract}
In this paper, we further investigate the problem of selecting a set of design points for universal kriging, which is a widely used technique for spatial data analysis. Our goal is to select the design points in order to make simultaneous predictions of the random variable of interest at a finite number of unsampled locations with maximum precision. Specifically, we consider as response  a correlated random field given by a linear model with an unknown parameter vector and a spatial error correlation structure. We propose a new design criterion that aims at simultaneously minimizing the variation of the prediction errors at various points. We also present various efficient techniques for incrementally building designs for that criterion scaling well for high dimensions. Thus the method is particularly suitable for big data applications in areas of spatial data analysis such as mining, hydrogeology, natural resource monitoring, and environmental sciences or equivalently for any computer simulation experiments. We have demonstrated the effectiveness of the proposed designs through two illustrative examples: one by simulation and another based on real data from Upper Austria.
\end{abstract}


\end{frontmatter}
\newdefinition{defin}{Definition}
\newdefinition{rem}{Remark}
\newtheorem{thm}{Theorem}
\newenvironment{myproof}[2] {\textsc{Proof of {#1} {#2}}:}

\newtheorem{conj}[thm]{Conjecture}
\newtheorem{coro}[thm]{Corollary}

\section{Introduction}
\label{intro}

Given a finite number $k+m$ of locations $x$ we are interested into making simultaneous predictions $\hat Y(\cdot)$ of $Y(\cdot)$ at $m$ unsampled locations using observations $Y(x_1),\ldots,Y(x_k)$ collected at some design points $\xi=(x_1,\ldots,x_k)\subset\SX^k$. Our objective is to select $\xi$ (of given size $k$) in order to maximize the precision of the predictions $\hat Y(x)$ over $\SX$. This setup is used in such diverse areas of spatial data analysis as mining, hydrogeology, natural resource monitoring and environmental sciences, see, \emph{e.g.}, \cite{cressie_93}, and has become the standard modeling paradigm in computer simulation experiments (cf. \cite{Fang+al_05,Kleijnen_09, rasmussen+w_05,  santner+al_03}), known under the designations of Gaussian Process (GP) modelling and kriging analysis. For a general review in the context of spatial statistics see \cite{wang_review_2012}.

Specifically, the model underlying our investigations is the model for universal kriging, i.e. we have a correlated scalar random
field given by
\begin{equation}\label{linmod}
  Y\left(x\right) = \mu(x,\boldsymbol{\beta})+\varepsilon(x)
\end{equation}

Here, $\boldsymbol{\beta}$ is an unknown vector of parameters in $\mathbb{R}^p$, $\mu(\cdot,\cdot)$ a known function of regressors at some given locations $x$ in a compact subset $\SX$ of $\mathbb{R}^d$ and the random term $\varepsilon \left( x\right)$ has zero mean, variance $\sigma^2$ and a parameterized spatial error correlation structure such that $\Ex\left(\varepsilon \left( x\right)\varepsilon \left( x'\right)\right)=\sigma^2 c(x,x';\nu)$ with $\nu$ some covariance parameters. We further assume that
\begin{itemize}
  \item the deterministic term is linear in the parameters $\boldsymbol{\beta}$, \emph{i.e.}, $\mu(x,\boldsymbol{\beta})=\textbf{f}(x)\boldsymbol{\beta}$, where\\ $\textbf{f}(x)=\left(f_1(x)\;f_2(x)\;\cdots\;f_p(x)\right)$ is a vector of known functions,
  \item the first two moments of the error $\varepsilon\left(x\right)$ and hence of $Y\left(x\right)$ exist and
  \item the variance $\sigma^2$ and the covariance parameters $\nu$ are known.
\end{itemize}
It is often assumed that the random field $\varepsilon \left( x\right)$ is Gaussian, allowing estimation of $\beta$ and $\theta=\{\sigma^2, \nu\}$ by Maximum Likelihood. We do not need to assume a stationary nor an isotropic covariance structure.

Traditional optimality criteria for designs for prediction are kriging variance-minimizing, albeit not all authors understand the same by the term \emph{kriging variance} or \emph{kriging covariance}. Some mean the variance of the prediction $\Var\left(\hat{Y}(x)\right)$,  cf. (\cite{Mueller+al_15}),
%\textbf{[cite someone]}
some the variance of the prediction error $\Var\left(\hat{Y}(x)-Y(x)\right)$, cf. (\cite{cressie_93}),
%\textbf{[cite someone]}
and some even $\Var\left(Y(x)|Y(\xi)\right)$, i.e. the variance of $Y(\cdot)$ at unsampled locations $x$ given the observations on the design points $\xi$,  cf. (\cite{Chevalier_Ginsbourger_12}).
We follow the second perception because trying to minimize the variation of the prediction errors seems to yield most precise predictions.

\begin{defin}
Let $x\in\SX$ be an arbitrary unsampled location and $\hat{Y}(x)$ the best linear unbiased predictor (BLUP) at $x$. The \emph{kriging variance} at $x$ is
$$
\sigma^2(x):=\Var\left(\hat{Y}(x)-Y(x)\right)
$$
\emph{i.e.} the variance of the best linear predictor minus the random variable to be predicted.

Let $x'\neq x$ be a second unsampled location and $\hat{Y}(x')$ the BLUP at $x'\in\SX$. The \emph{kriging covariance} for $x$ and $x'$ is
$$
\sigma(x,x'):=\Cov\left(\hat{Y}(x)-Y(x);\hat{Y}(x')-Y(x')\right)
$$
\end{defin}
\smallskip

\noindent With the above definition G-optimal designs (cf. \cite{dasgupta_g-optimal_2022}) try to minimize the maximum kriging variance, \emph{i.e.}
\begin{equation} \label{Gopt}
\min_{\xi} \max_{x \in \SX}  \Var\left(\hat Y(x)-Y(x)\right)\, .
\end{equation}
Another popular optimality criterion tries to minimize the average prediction variance over a set of $m$ specific points (V-optimality), \emph{i.e.}
\begin{equation} \label{Vopt}
\min_{\xi} \frac{1}{m}\sum_{x_i \in \SX}  \Var\left(\hat Y(x_i)-Y(x_i)\right)\, .
\end{equation}
Note that the latter if not supported on a grid but rather covering the whole $\SX$, expressed as an integral is often called I-optimality, see \cite{dasgupta_optimal_2022-1} for a recent example of the terminology.

None of these and other criteria for designs for prediction considers the \emph{kriging covariances} or the \emph{kriging covariance matrix} at the unsampled locations $(x_{k+1},\ldots,x_{k+m})$
\[\boldsymbol\Sigma=\boldsymbol\Sigma(x_{k+1},\ldots,x_{k+m})=\left(
                                \begin{array}{cccc}
                                  \sigma^2(x_{k+1}) & \sigma(x_{k+1},x_{k+2}) & \cdots & \sigma(x_{k+1},x_{k+m}) \\
                                  \sigma(x_{k+1},x_{k+2}) & \sigma^2(x_{k+2}) & \cdots & \sigma(x_{k+2},x_{k+m}) \\
                                  \vdots & \vdots & \ddots & \vdots \\
                                  \sigma(x_{k+1},x_{k+m}) & \sigma(x_{k+2},x_{k+m}) & \cdots & \sigma^2(x_{k+m}) \\
                                \end{array}
                              \right)
\]
The popular design criteria just use the diagonal of $\boldsymbol\Sigma$ which means to voluntarily dispense of valuable information given by the kriging covariances.
\smallskip

The design criterion considered in this paper, is the \emph{generalized variance} of the kriging covariance matrix:
\begin{equation}\label{GVopt}
GV(\xi) = \det\boldsymbol\Sigma(x_{k+1},\ldots,x_{k+m})=|\boldsymbol\Sigma|\,.
\end{equation}

\begin{defin}
The design $\xi$ that minimizes criterion (\ref{GVopt}) or equivalently any root of ${|\boldsymbol\Sigma|}$ is defined as the GV-optimal design, GV stands for Generalized (kriging) Variance.
\end{defin}
\smallskip

\begin{figure}
\begin{center}
\includegraphics[width=\textwidth]{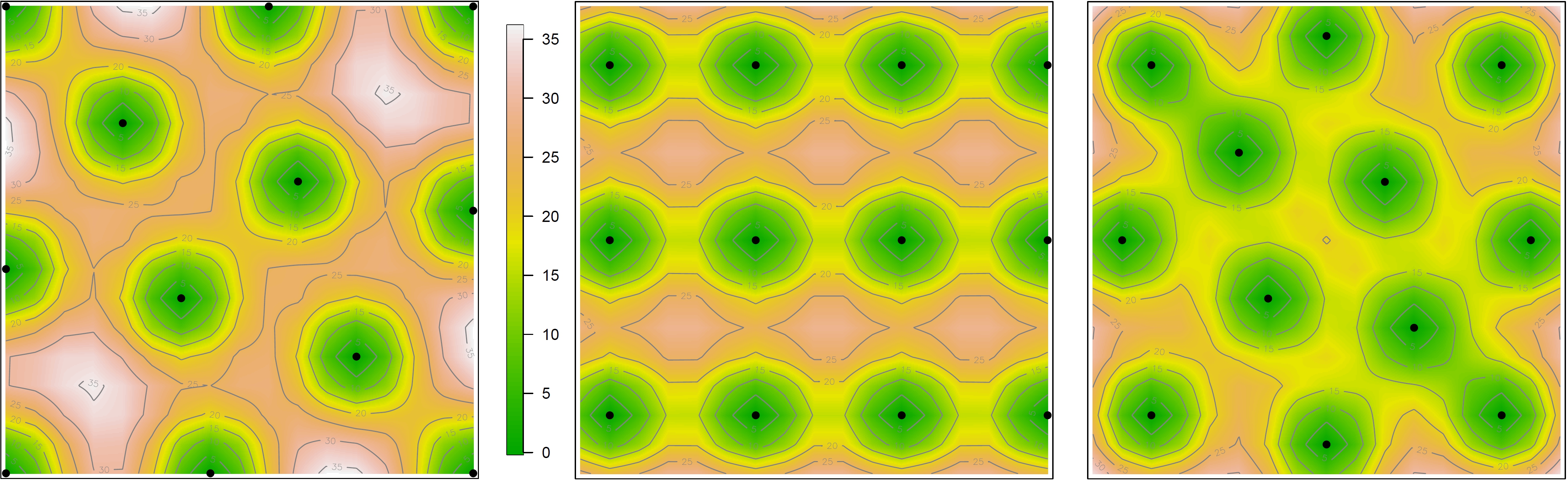}
\end{center}
\caption{\footnotesize Map of kriging variances of GV-optimal (left), G-optimal (middle) and V-optimal (right) 12-point design for the ordinary kriging setup on the unit square and Matern covariance function with $\kappa=2.5$ and $\phi=3$. The black dots are the design points.
Note that in some zones of the design region the GV-optimal design has considerable larger kriging variances than the G- and V-optimal designs. This is a natural consequence of the GV-criterion, which does not aim at minimizing the point-wise kriging variances while the G- and V- criterion do.
%On the other hand the figure shows that GV-optimal designs also perform well with respect to the kriging variances compared to the variance minimizing counterparts.
}
\label{F:compareOK}
\end{figure}

As designs that simply minimize the kriging variance, GV-optimal designs are often space-filling but typically position more design points at the edge of the design region. GV-optimal designs depend more on the special covariance structure which is in contrast to G- and V-optimal designs in particular for small numbers of observations $k$ (see Figure \ref{F:compareOK} for an exemplary comparison). Unfortunately the  minimization of the GV-criterion  is computationally  demanding, since the evaluation of (\ref{GVopt}) requires the evaluation of the determinant of an $(m\times m)$-matrix, being unfeasible for large $m$ which is especially necessary in high dimensional design spaces as it is often the case for computer experiments. A remedy for this problem is provided by the use of incrementally assembled designs proposed here which turn out to be GV-optimal. This reduces the computational effort for the evaluation of (\ref{GVopt}) for arbitrary $m$ to the evaluation of the determinant of a $(k\times k)$-matrix, where $k$ is just the design size.

The paper is organized as follows. In Section \ref{S:GV} we motivate our approach, exploiting the fact that the volume of the simultaneous confidence region for the prediction errors is proportional to $\sqrt{|\boldsymbol\Sigma|}$. Section \ref{S:INCR} presents the basis of the main contribution of the paper, giving an update formula for the determinant of the kriging covariance matrix which should be used if we are constructing our designs incrementally. These update formulas are also necessary to speed up the computation of the design criterion (\ref{GVopt}) and may as well be used for a more efficient computation of the design criterion V-optimal designs. Finally, Section \ref{S:EFF} considers the efficiency of incremental GV-optimal designs and also the efficiency of GV-optimal designs with respect to G- and V-optimal designs which is demonstrated by means of a representative simulation study in Section \ref{S:EX}, followed by a real world example in Section \ref{S:RWEX}, and Section \ref{S:CONCL} draws conclusions on the efficiency and limitations of the approach and suggests topics for future work.

%_________________________________________________________________________________________________________________
\section{The generalized variance and optimal designs for prediction}\label{S:GV}

\cite{Wilks_32} remarked that there were "\ldots statistical coefficients which have not been adequately generalized for samples from a multivariate population including the variance \ldots" and introduced the \emph{generalized variance} which is simply the determinant of the covariance matrix of a multivariate population.
%Abraham
\cite{Wald_43} used the idea of minimizing the generalized variance in his criterion for optimal designs for parameter estimation (D-optimality), \emph{i.e.}
$$
\max_{\xi} |M_{\theta}(\xi)|\, ,
$$
where $M_{\theta}$ is the information matrix of the parameter estimates $\hat{\theta}$. A combination of these two ideas naturally leads to criterion (\ref{GVopt}).

\cite{Shewry_Wynn_87} introduced maximum entropy sampling (MES) where the Shannon entropy is used as a measure of information to get optimal designs for prediction. Here the design criterion is similar to (\ref{GVopt}): in the case of Gaussian response $Y$, MES tries to maximize the determinant of the covariance matrix of $Y(\xi)$ which is equivalent to minimizing the determinant of the covariance matrix of $\left(Y(x_{k+1}),\ldots,Y(x_{k+m})|Y(x_{1}),\ldots,Y(x_{k})\right)$, \emph{i.e.} the determinant of the covariance matrix of $Y(\cdot)$ at the unsampled locations $(x_{k+1},\ldots,x_{k+m})$ conditioned on $Y(\cdot)$ at sampled locations $(x_{1},\ldots,x_{k})$. Though this criterion does not aim at minimizing the variation of the prediction errors, it is not even directly connected with a prediction method. MES it is rather a sampling method trying to absorb the maximum amount of variability into the sample, such that conditional on the sample the unsampled points have minimum variability. The method is suitable for observations on a finite closed system which is the main connection to the present work. Beyond that we connect the sampling method directly with the BLUP for the response $Y$ as presented hereinafter.

Interestingly MES yields designs that are exactly GV-optimal in the simple kriging setup and also seem to be GV-optimal in the ordinary kriging setup. For universal kriging GV-optimal designs approach MES designs with increasing effective range, \emph{i.e.} more strongly correlated response.

%---------------------------------------------------------------------------------------------------
\subsection{The best linear unbiased predictor and the corresponding kriging covariance matrix}\label{S:BLUP}

As with MES we assume that the allowable choice of the designs $\xi$ is from a fixed finite set of $N=k+m$ points $X=\left\{x_1,\ldots,x_{k+m}\right\}$. Given a $k$-point design $\xi=(x_1,\ldots,x_k)$, $0<k<N$, the complementary design $\xi_0$ is the set $X\setminus\xi$ and we get the corresponding partitioning of the response vector $\mathbf{Y}=\left(\mathbf{Y}_{\xi}^T\;\mathbf{Y}_0^T\right)^T$. We now want to simultaneously predict the response from the complementary design on the basis of the response from the design $\xi$.

As mentioned above we are using the model of universal kriging, i.e. our linear predictor is the linear combination of a vector of deterministic functions of the locations $x\in\xi$: $\textbf{f}(x)=\left(f_1(x)\;f_2(x)\;\cdots\;f_p(x)\right)$ with the first component usually being $f_1(x)=1$. The design matrix and corresponding vector of errors then are
\[\mathbf{F}_{\xi}=\left(
                   \begin{array}{cccc}
                     f_1(x_1) & f_2(x_1) & \ldots & f_p(x_1) \\
                     f_1(x_2) & f_2(x_2) & \ldots & f_p(x_2) \\
                     \vdots & \vdots & \vdots & \vdots \\
                     f_1(x_k) & f_2(x_k) & \ldots & f_p(x_k) \\
                   \end{array}
                 \right)\qquad\boldsymbol{\varepsilon}_{\xi}=\left(
                                                               \begin{array}{c}
                                                                 \varepsilon(x_1) \\
                                                                 \varepsilon(x_2) \\
                                                                 \vdots \\
                                                                 \varepsilon(x_k) \\
                                                               \end{array}
                                                             \right)
\]
We denote the covariance matrix of $\boldsymbol{\varepsilon}_{\xi}$ with $\mathbf{C}_{\xi}$ and use analogous nomenclature for the complementary design to get
\[\Ex\left(
    \begin{array}{c}
      \mathbf{Y}_{\xi} \\
      \mathbf{Y}_0 \\
    \end{array}
  \right)=\left(
            \begin{array}{c}
            \mathbf{F}_{\xi} \\
            \mathbf{F}_0 \\
            \end{array}
          \right)\boldsymbol{\beta}\qquad\qquad\Cov\left(
                                                    \begin{array}{c}
                                                    \mathbf{Y}_{\xi} \\
                                                    \mathbf{Y}_0 \\
                                                    \end{array}
                                                \right)=\left(
                                                     \begin{array}{cc}
                                                       \mathbf{C}_{\xi} & \mathbf{C}_{\xi 0} \\
                                                       \mathbf{C}_{\xi 0}^T & \mathbf{C}_{0} \\
                                                     \end{array}
                                                   \right)\qquad\Rightarrow
\]
\[
  \Rightarrow\qquad\Ex\left(\mathbf{Y}_0|\mathbf{Y}_{\xi}\right)=\mathbf{F}_0\boldsymbol{\beta}+\mathbf{C}_{\xi 0}^T\mathbf{C}_{\xi}^{-1}\left(\mathbf{Y}_{\xi}-\mathbf{F}_{\xi}\boldsymbol{\beta}\right)\qquad\qquad \Cov\left(\mathbf{Y}_0|\mathbf{Y}_{\xi}\right)=\mathbf{C}_{0}-\mathbf{C}_{\xi 0}^T\mathbf{C}_{\xi}^{-1}\mathbf{C}_{\xi 0}
\]
The generalized least squares (GLS) estimate of the parameter vector then is $\hat{\boldsymbol{\beta}}=\left(\mathbf{F}_{\xi}^T\mathbf{C}_{\xi}^{-1}\mathbf{F}_{\xi}\right)^{-1} \mathbf{F}_{\xi}^T\mathbf{C}_{\xi}^{-1}\mathbf{Y}_{\xi}$ yielding the simultaneous kriging prediction for all complementary design points as the BLUP
\[\hat{\mathbf{Y}}_{0}=\mathbf{F}_0\hat{\boldsymbol{\beta}}+\mathbf{C}_{\xi 0}^T\mathbf{C}_{\xi}^{-1}\left(\mathbf{Y}_{\xi}-\mathbf{F}_{\xi}\hat{\boldsymbol{\beta}}\right)= \left(\mathbf{F}_0\mathbf{B}+\mathbf{C}_{\xi 0}^T\mathbf{A}\right)\mathbf{Y}_{\xi} = \mathbf{W}\mathbf{Y}_{\xi}
\]
where $\mathbf{B}=\left(\mathbf{F}_{\xi}^T\mathbf{C}_{\xi}^{-1}\mathbf{F}_{\xi}\right)^{-1} \mathbf{F}_{\xi}^T\mathbf{C}_{\xi}^{-1}$ and $\mathbf{A}=\mathbf{C}_{\xi}^{-1}\left(\mathbf{I}_k-\mathbf{F}_{\xi}\mathbf{B}\right)$. The components of the $(m\times k)-$\emph{weight matrix} $\mathbf{W}=\left(\omega_{ij}\right)$ give the kriging weights of $Y(x_j)$ for the prediction $\hat{Y}(x_{k+i})$. Combining the results from above the kriging prediction errors have expectation zero and the \emph{kriging covariance matrix}
\begin{equation}\label{krigingcov}
  \Cov\left(\hat{\mathbf{Y}}_{0}-\mathbf{Y}_{0}\right)=\boldsymbol{\Sigma}=\mathbf{W}\mathbf{C}_{\xi}\mathbf{W}^{T}-\mathbf{W}\mathbf{C}_{\xi 0}-\mathbf{C}_{\xi 0}^T\mathbf{W}^T+\mathbf{C}_{0}
\end{equation}

%---------------------------------------------------------------------------------------------------
\subsection{Geometrical interpretation of the GV criterion}\label{S:GEOMINT}

%As we assumed the existence of the first two moments of $Y(x)$ the kriging prediction errors are asymptotically normal
%\[\left(\hat{\mathbf{Y}}_{0}-\mathbf{Y}_{0}\right)\stackrel{d}{\rightarrow}\N\left(\,\mathbf{0}\,;\,\boldsymbol{\Sigma}\,\right)\qquad\Rightarrow\qquad %\left(\hat{\mathbf{Y}}_{0}-\mathbf{Y}_{0}\right)^T\boldsymbol{\Sigma}\left(\hat{\mathbf{Y}}_{0}-\mathbf{Y}_{0}\right)\stackrel{d}{\rightarrow} \boldsymbol{\chi}^2_m
%\]
%We get an asymptotic simultaneous confidence region for the prediction errors with

In the case of Gaussian kriging prediction errors,
\[\left(\hat{\mathbf{Y}}_{0}-\mathbf{Y}_{0}\right)\sim\N\left(\,\mathbf{0}\,;\,\boldsymbol{\Sigma}\,\right)\qquad\Rightarrow\qquad \left(\hat{\mathbf{Y}}_{0}-\mathbf{Y}_{0}\right)^T\boldsymbol{\Sigma}^{-1}\left(\hat{\mathbf{Y}}_{0}-\mathbf{Y}_{0}\right)\sim \boldsymbol{\chi}^2_m
\]
we get a simultaneous confidence region for the prediction errors with
\[\left\{\hat{\mathbf{Y}}_{0}-\mathbf{Y}_{0}:\left(\hat{\mathbf{Y}}_{0}-\mathbf{Y}_{0}\right)^T\boldsymbol{\Sigma}^{-1}\left(\hat{\mathbf{Y}}_{0}-\mathbf{Y}_{0}\right) \leq\boldsymbol{\chi}^2_{m;1-\alpha}\right\}
\]
where $\boldsymbol{\chi}^2_{m;1-\alpha}$ is the $(1-\alpha)$-quantile of the $\boldsymbol{\chi}^2_m$ distribution.

This region is a $k$-dimensional ellipsoid who's volume is proportional to $\sqrt{|\boldsymbol\Sigma|}$, \emph{i.e.} GV-optimal designs minimize the area of the prediction errors. It is evident that the usually used design criteria like G-optimality or V-optimality do not have this highly desirable property.

%---------------------------------------------------------------------------------------------------
\subsection{Comparison of MES and the GV-criterion}

The criterion for a GV-optimal design or a GV-optimal increment (\ref{updateGVcriterion}) looks similar to the criterion of MES with multivariate normal distributed $\mathbf{Y}$, which in our notation would be to maximize $\left|\Cov\left(\mathbf{Y}_{\xi}\right)\right|=\left|\mathbf{C}_{\xi}\right|$, which is equivalent to minimizing $\left|\Cov\left(\mathbf{Y}_0|\mathbf{Y}_{\xi}\right)\right|=\left|\mathbf{C}_0-\mathbf{C}_{\xi 0}^T\mathbf{C}_{\xi}^{-1}\mathbf{C}_{\xi 0}\right|$. The reason for the equivalence lies in the determinant identity $\left|\left(
                                                     \begin{array}{cc}
                                                       \mathbf{C}_{\xi} & \mathbf{C}_{\xi 0} \\
                                                       \mathbf{C}_{\xi 0}^T & \mathbf{C}_{0} \\
                                                     \end{array}
                                                   \right)\right|=\left|\mathbf{C}_{\xi}\right|\cdot\left|\mathbf{C}_0-\mathbf{C}^T_{\xi 0}\mathbf{C}^{-1}_{\xi}\mathbf{C}_{\xi 0}\right|$ and the fact that the left hand side is fixed and finite (see \cite{Shewry_Wynn_87}).

However, there are considerable differences between the GV-criterion and the MES-criterion:
\begin{itemize}
  \item The above MES-criterion just holds in the case of Gaussian $\mathbf{Y}$. We do not assume multivariate normal distributed $\mathbf{Y}$ for our GV-criterion, \emph{i.e.} the two criteria are different in the case of non-Gaussian $\mathbf{Y}$.
  \item With GV-optimal designs we minimize $\Cov\left(\hat{\mathbf{Y}}_{0}-\mathbf{Y}_0\right)$ in contrast to $\Cov\left(\mathbf{Y}_0|\mathbf{Y}_{\xi}\right)$ which is minimized with MES-optimal designs. This difference just vanishes in the case of simple kriging which is shown later.
  \item The implementation of MES for a linear model (\ref{linmod}) uses a Bayesian model which needs some prior distribution for the parameter vector $\boldsymbol{\beta}$ and particularly some prior covariance matrix $\Cov\left(\boldsymbol{\beta}\right)=\mathbf{R}^{-1}$. The MES criterion for a linear model then turns out to be: maximize $\left|\mathbf{C}_{\xi}+\mathbf{F}_{\xi}\mathbf{R}^{-1}\mathbf{F}_{\xi}^T\right|$. We have
      \[\left|\mathbf{C}_{\xi}+\mathbf{F}_{\xi}\mathbf{R}^{-1}\mathbf{F}_{\xi}^T\right|= \left|\mathbf{C}_{\xi}\right|\left|\mathbf{I}_{p}+\mathbf{R}^{-\frac{1}{2}}\mathbf{F}_{\xi}^T\mathbf{C}_{\xi}^{-1}\mathbf{F}_{\xi}\mathbf{R}^{-\frac{1}{2}}\right|= \left|\mathbf{C}_{\xi}\right|\left|\mathbf{R}^{-1}\right|\left|\mathbf{R}+\mathbf{F}_{\xi}^T\mathbf{C}_{\xi}^{-1}\mathbf{F}_{\xi}\right|
      \]
      and since $\mathbf{R}$ does not depend on the design $\xi$ the MES criterion for a linear model can also be formulated as: Select $\xi$ such that $\left|\mathbf{C}_{\xi}\right|\left|\mathbf{R}+\mathbf{F}_{\xi}^T\mathbf{C}_{\xi}^{-1}\mathbf{F}_{\xi}\right|$ is maximized. Interestingly maximizing $\left|\mathbf{R}+\mathbf{F}_{\xi}^T\mathbf{C}_{\xi}^{-1}\mathbf{F}_{\xi}\right|$ alone is known as Bayesian D-optimality \cite{Chaloner_Verdinelli_95}.

      We are not working in a Bayesian framework and do not assume a prior distribution or a prior covariance matrix for $\boldsymbol\beta$, so the GV-criterion is clearly different from MES for a linear model.
\end{itemize}
Despite these differences it might be interesting to compare GV-optimal designs with MES for Gaussian response just using a constant model. Here the MES-criterion is simply: choose $\xi$ such that $\left|\mathbf{C}_{\xi}\right|$ is maximized. For simple kriging this criterion is equivalent to the GV-criterion  because here the kriging covariance matrix is $\Cov\left(\hat{\mathbf{Y}}_{0}-\mathbf{Y}_0\right)=\mathbf{C}_0-\mathbf{C}_{\xi 0}^T\mathbf{C}_{\xi}^{-1}\mathbf{C}_{\xi 0}$ and minimizing $\left|\boldsymbol\Sigma_{\textrm{SK}}\right|=\left|\mathbf{C}_0-\mathbf{C}_{\xi 0}^T\mathbf{C}_{\xi}^{-1}\mathbf{C}_{\xi 0}\right|$ is equivalent to maximizing $\left|\mathbf{C}_{\xi}\right|$ (see above).
\medskip

\begin{figure}
\begin{center}
\includegraphics[width=.5\textwidth]{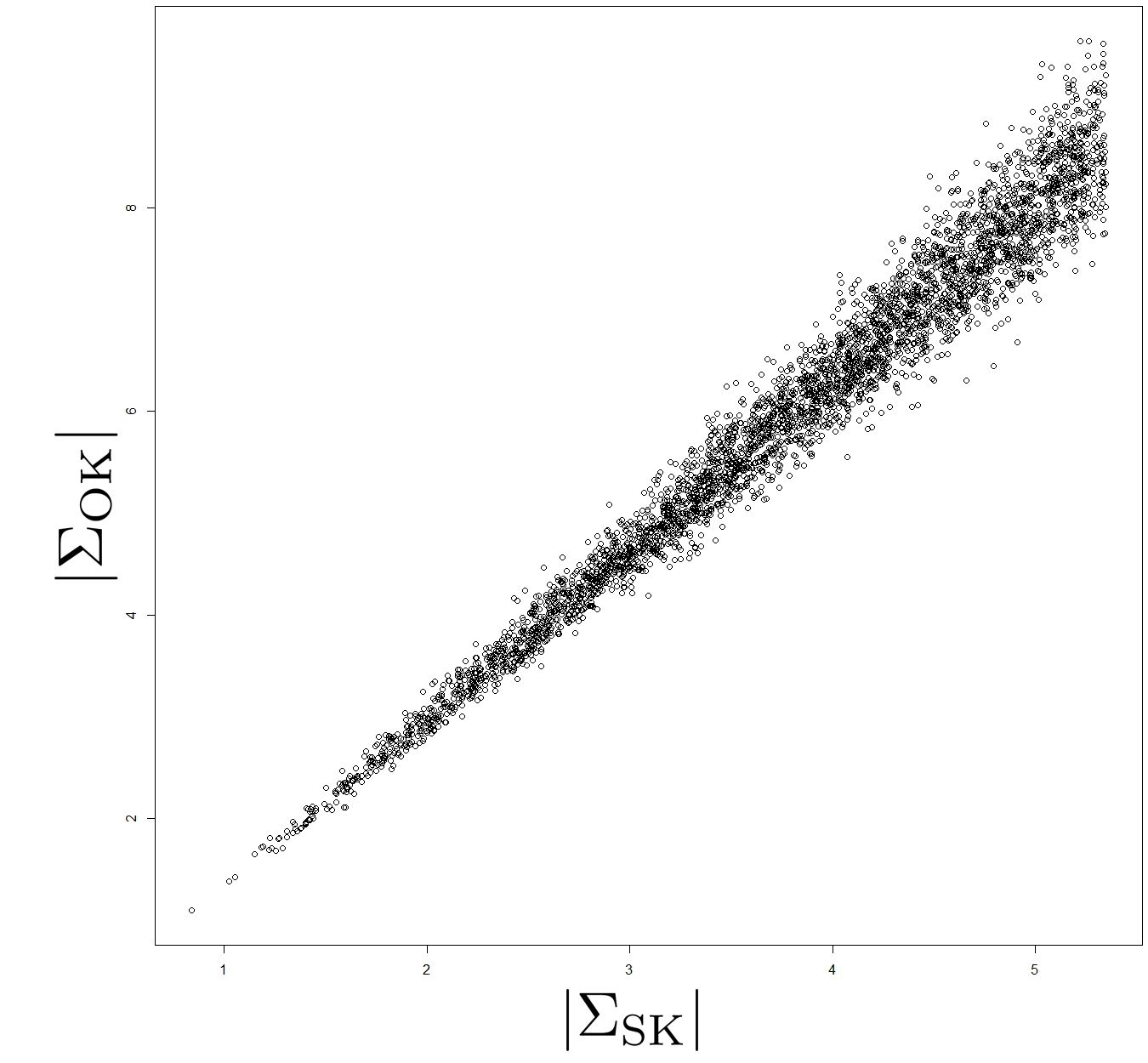}
\end{center}
\caption{\footnotesize GV-criterion function for simple kriging $\left|\boldsymbol\Sigma_{\textrm{SK}}\right|$ against GV-criterion function for ordinary kriging $\left|\boldsymbol\Sigma_{\textrm{OK}}\right|$ for different designs. The minima of both criteria seem to be the same.}
\label{F:compareSKOK}
\end{figure}

With ordinary kriging the above equivalence is not clear: The kriging weights for ordinary kriging are
\[\mathbf{W}=\frac{1}{b}\mathds{1}_m\mathds{1}_k^T\mathbf{C}_{\xi}^{-1}+ \mathbf{C}_{\xi 0}^T\mathbf{C}_{\xi}^{-1}\left(\mathbf{I}_k-\frac{1}{b}\mathds{1}_m\mathds{1}_k^T\mathbf{C}_{\xi}^{-1}\right)
\]
where $b=\mathds{1}_k\mathbf{C}_{\xi}^{-1}\mathds{1}_k^T$ and $\mathds{1}_n$ is a vector of ones with length $n$. The kriging covariance matrix for ordinary kriging then is
\[\boldsymbol\Sigma_{\textrm{OK}}=\mathbf{C}_0-\mathbf{C}_{\xi 0}^T\mathbf{C}_{\xi}^{-1}\mathbf{C}_{\xi 0}+\frac{1}{b}\left(\mathds{1}_m-\mathbf{C}_{\xi 0}^T\mathbf{C}_{\xi}^{-1}\right)\left(\mathds{1}_m-\mathbf{C}_{\xi 0}^T\mathbf{C}_{\xi}^{-1}\right)^T
\]
and the GV-criterion in this case requires the minimizing of
\[\left|\boldsymbol\Sigma_{\textrm{OK}}\right|=\left|\mathbf{C}_0-\mathbf{C}_{\xi 0}^T\mathbf{C}_{\xi}^{-1}\mathbf{C}_{\xi 0}\right|\textrm{abs}\left(1+\frac{1}{b}\left(\mathds{1}_m-\mathbf{C}_{\xi 0}^T\mathbf{C}_{\xi}^{-1}\right)^T\left(\mathbf{C}_0-\mathbf{C}_{\xi 0}^T\mathbf{C}_{\xi}^{-1}\mathbf{C}_{\xi 0}\right)^{-1}\left(\mathds{1}_m-\mathbf{C}_{\xi 0}^T\mathbf{C}_{\xi}^{-1}\right)\right)
\]
The above criterion function is clearly different from the simple kriging setup and the following conjecture was verified by countless computations of $\left|\boldsymbol\Sigma_{\textrm{SK}}\right|$ and $\left|\boldsymbol\Sigma_{\textrm{OK}}\right|$ for the same randomly generated designs respectively with different covariance models, different functions $\mathbf{f}(x)$ and designs of different sizes. A formal proof eludes us.

\begin{conj}
Let $\left|\boldsymbol\Sigma_{\textrm{SK}}^1\right|$ be the GV-criterion function for design $\xi_1$ in the simple kriging setup and $\left|\boldsymbol\Sigma_{\textrm{OK}}^1\right|$ the criterion function in the ordinary kriging setup for the same design $\xi_1$. $\left|\boldsymbol\Sigma_{\textrm{SK}}^2\right|$ and $\left|\boldsymbol\Sigma_{\textrm{OK}}^2\right|$ are defined analogously for another design $\xi_2$. Then
\[\left|\boldsymbol\Sigma_{\textrm{SK}}^1\right|\lesseqgtr\left|\boldsymbol\Sigma_{\textrm{SK}}^2\right|\qquad\nLeftrightarrow\qquad \left|\boldsymbol\Sigma_{\textrm{OK}}^1\right|\lesseqgtr\left|\boldsymbol\Sigma_{\textrm{OK}}^2\right|,
\]
but the arguments of the minima are the same in both cases as can be seen in Figure \ref{F:compareSKOK}.
\[\arg\min_{\xi}\left|\boldsymbol\Sigma_{\textrm{SK}}\right|\;=\;\arg\min_{\xi}\left|\boldsymbol\Sigma_{\textrm{OK}}\right|
\]
\end{conj}
\medskip

In the case of universal kriging the design criterium and the optimal design is clearly different from MES even though it can be observed that with increasing effective range \emph{i.e.} with higher correlated response GV-optimal designs approach MES-designs also for universal kriging.

%_________________________________________________________________________________________________________________
\section{Incrementally and decrementally constructed \\GV-optimal designs}\label{S:INCR}

In many practical situations the experiment is not stopped after a fixed number of runs but say
%\emph{e.g.}
after a certain time or when a certain budget for the runs has expired. Thus at the start of the experiment the sample size is unknown and it is not clear which optimal designs of which size to use. In these situation incremental designs should be used, starting with a first design of (small) size $k$ and supplementing it step by step with increments of size $l_i\geq 1$ until experimenting has to be stopped.
Here the question arises how to find optimal increments in an efficient way which will be answered in this section. The efficiency of incrementally built designs compared to GV-optimal designs will be examined in Section \ref{S:EFF}.

The use of decrementally constructed designs is not directly motivable, it arises from the attempt of detecting GV-optimal designs of size $k+l$ in an incremental way. This may be done very efficiently with the help of Corollary 1 as will be shown below. The idea is to start with a design of minimal size $k$ and compute an increment of size $l$ which may be done with small computational effort. The only problem can be caused by the starting design which may contain design points which are not elements of the GV-optimal design of size $k+l$ leading to highly efficient but still suboptimal designs. In these situations an improvement can be achieved using a decrement, i.e. omitting $k_1$ design points followed by another incremental step. The hope in the decremental step is to get rid of inappropriate design points which in the following incremental step are substituted by design points yielding designs closer to GV-optimality.

\subsection{Update formulae for kriging weights and the kriging covariance matrix of incremental designs}\label{updateincrement}

For this subsection we assume that we already have a $k$-point design and want to add $l$ extra design points to simultaneously predict the remaining $m\gg l$ non-design points. As the calculation of the design criterion is computationally demanding, we will show that the use of update formulae for the kriging weights and consequently for the kriging covariance matrix is of great computational benefit.

Furthermore these update formulae may also be used for updating $\Cov\left(\hat{Y}(x)\right)$ and $\Cov\left(Y(x)|Y(\xi)\right)$ and also for incrementally built G- and V-optimal designs.
\medskip

The allowable choice of the designs $\xi_1$ and $\xi_2$ with $\xi_1\cap\xi_2=\emptyset$ are from a fixed finite set of $N=k+l+m$ points $X=\{x_1,\ldots,x_{k+l+m}\}$. $\xi_1=\{x_1,\ldots,x_{k}\}$ is the $k$-point first or starting design, $\xi_2=\{x_{k+1},\ldots,x_{k+l}\}$ is the $l$-point second design (the increment), and  $X\setminus\{\xi_1\cup\xi_2\}$ is the remaining sets of non-design points with cardinality $m$. Thus we get the corresponding partitioning of the response vector $\mathbf{Y}=\left(\mathbf{Y}_{\xi_1}^T\;\mathbf{Y}_{\xi_2}^T\;\mathbf{Y}_0^T\right)^T$ or simpler $\mathbf{Y}=\left(\mathbf{Y}_1^T\;\mathbf{Y}_2^T\;\mathbf{Y}_0^T\right)^T$ with expectation and covariance matrix
\begin{equation}\label{partitionedmod}
\Ex\left(
    \begin{array}{c}
      \mathbf{Y}_1 \\
      \mathbf{Y}_2 \\
      \mathbf{Y}_0 \\
    \end{array}
  \right)=\left(
            \begin{array}{c}
            \mathbf{F}_1 \\
            \mathbf{F}_2 \\
            \mathbf{F}_0 \\
            \end{array}
          \right)\boldsymbol{\beta}\qquad\qquad\Cov\left(
                                                    \begin{array}{c}
                                                    \mathbf{Y}_1 \\
                                                    \mathbf{Y}_2 \\
                                                    \mathbf{Y}_0 \\
                                                    \end{array}
                                                \right)=\left(
                                                     \begin{array}{ccc}
                                                       \mathbf{C}_1 & \mathbf{C}_{1\, 2} & \mathbf{C}_{1\, 0} \\
                                                       \mathbf{C}_{1\, 2}^T & \mathbf{C}_2 & \mathbf{C}_{2\, 0} \\
                                                       \mathbf{C}_{1\, 0}^T & \mathbf{C}_{2\, 0}^T & \mathbf{C}_0 \\
                                                     \end{array}
                                                   \right).
\end{equation}
In the first stage the design is $\xi_1$ and we have to predict $\left(\mathbf{Y}_2^T\;\mathbf{Y}_0^T\right)^T$ from $\mathbf{Y}_1$:
\begin{equation}\label{stage1}
\left(
    \begin{array}{c}
      \hat{\mathbf{Y}_2} \\
      \hat{\mathbf{Y}_0} \\
    \end{array}
  \right)=\left(
            \begin{array}{c}
              \mathbf{W}_{12} \\
              \mathbf{W}_{10} \\
            \end{array}
          \right)\mathbf{Y}_1,
\end{equation}
where $\mathbf{W}_{12}$ are the weights of $\mathbf{Y}_1$ for the prediction of $\mathbf{Y}_2$ and analogously $\mathbf{W}_{10}$ the weights of $\mathbf{Y}_1$ for the prediction of $\mathbf{Y}_0$.

In the second stage we add the increment and the design is now $\xi_1\cup\xi_2$ and we have to predict $\mathbf{Y}_0$ from $\left(\mathbf{Y}_1^T\,\mathbf{Y}_2^T\right)^T$ as
\begin{equation}\label{stage2}
\hat{\mathbf{Y}_0}=\left(
            \begin{array}{cc}
              \mathbf{W}_{1} & \mathbf{W}_{2} \\
            \end{array}
          \right)\left(
                   \begin{array}{c}
                     \mathbf{Y}_1 \\
                     \mathbf{Y}_2 \\
                   \end{array}
                 \right)
\end{equation}
where $\mathbf{W}_{1}$ are the new weights of $\mathbf{Y}_1$ for the prediction of $\mathbf{Y}_0$ and analogously $\mathbf{W}_{2}$ the weights of $\mathbf{Y}_2$ for the prediction of $\mathbf{Y}_0$.

\cite{Emery_09} showed that the kriging weights of the first stage (in our notation the components of $\mathbf{W}_{12}$ and $\mathbf{W}_{10}$) can be updated such that in our compact matrix notation
\begin{equation}\label{updateW21}
\mathbf{W}_{1} = \mathbf{W}_{10}-\mathbf{W}_{2}\mathbf{W}_{12}\,.
\end{equation}
The only problem with (\ref{updateW21}) is that $\mathbf{W}_{2}$ is unknown if we add an increment to a smaller design. So, an "update" formula for the computation of the weights $\mathbf{W}_{2}$ (in fact this is not an update because weights of $\mathbf{Y}_2$ do not exist in the first stage) is essential for an efficient prediction of $\mathbf{Y}_0$ in the second stage.

\cite{Emery_09} also presented update formulae for the kriging variances and covariances which unfortunately are wrong in the case of $l>1$, which was shown with a simple counter example by \cite{Chevalier_Ginsbourger_12}. The presented ``corrected'' update formulae just have not been for the kriging covariance but for $\Cov\left(\mathbf{Y}_{0}|\mathbf{Y}_{\xi}\right)$. Eventually \cite{Chev_Gins_Emery_14} introduced correct update formulae for kriging variances and kriging covariances and also formulae for the new kriging weights of the second stage $\mathbf{W}_{2}$. In our notation these formulae may be summarized as follows.

\subsubsection{(Updated) kriging weights for the second stage}
Let the weights $\mathbf{W}_{12}$, $\mathbf{W}_{10}$, $\mathbf{W}_{1}$ and $\mathbf{W}_{2}$ be as defined in (\ref{stage1}) and (\ref{stage2}). Let further the kriging covariance matrix of the first stage in obvious notation be
\begin{equation}\label{krigcov1}
  \left(
    \begin{array}{cc}
      \boldsymbol\Sigma_2 & \boldsymbol\Sigma_{20} \\
      \boldsymbol\Sigma_{20}^T & \boldsymbol\Sigma_0 \\
    \end{array}
  \right)
\end{equation}
Then the new weights $\mathbf{W}_{1}$ and $\mathbf{W}_{2}$ can simply be computed with
\begin{eqnarray}\label{updateW22}
 \nonumber % Remove numbering (before each equation)
  \mathbf{W}_{1} &=& \mathbf{W}_{10}-\boldsymbol\Sigma_{20}^T\boldsymbol\Sigma_2^{-1}\mathbf{W}_{12} \\
  \mathbf{W}_{2} &=& \boldsymbol\Sigma_{20}^T\boldsymbol\Sigma_2^{-1}
\end{eqnarray}
An algebraic proof for the simultaneous computation of the $(m\times l)$-weight matrix $\mathbf{W}_2$ which is fundamentally different from the one given in \cite{Chev_Gins_Emery_14} can be found in the Appendix.

\subsubsection{Update formula for the kriging covariance matrix}
Let the kriging covariance matrix of the first stage be as in (\ref{krigcov1}), and the kriging covariance matrix of the second stage be $\boldsymbol\Sigma_0^+$.
Using the weights of (\ref{updateW22}) we can then update the kriging covariance matrix from the first stage to get
\begin{equation}\label{updatekrigcov}
  \boldsymbol\Sigma_0^+=\boldsymbol\Sigma_0-\boldsymbol\Sigma_{20}^T\boldsymbol\Sigma_2^{-1}\boldsymbol\Sigma_{20}
\end{equation}
An algebraic proof for this simultaneous update formula for the kriging covariance matrix which uses again another reasoning than \cite{Chev_Gins_Emery_14} can be found in the Appendix.
\medskip

\subsection{Efficient computation of increments for GV-optimal designs}

The following theorem is of fundamental importance as it allows the efficient incremental construction of $GV$-optimal designs.

\begin{thm}\label{thm2}
The GV-optimal increment for the second stage given a design at the first stage then is
\begin{equation}\label{updateGVcriterion}
  \xi_{\textrm{GV}}^{inc}=\arg\max_{\xi_2}\left|\boldsymbol\Sigma_2\right|.
\end{equation}
\end{thm}

\begin{myproof}{Theorem}{\ref{thm2}}
For a given design at the first stage the determinant of the corresponding kriging covariance matrix is fixed and finite and the determinant may be factored as:
\[\left|\left(
    \begin{array}{cc}
      \boldsymbol\Sigma_2 & \boldsymbol\Sigma_{20} \\
      \boldsymbol\Sigma_{20}^T & \boldsymbol\Sigma_0 \\
    \end{array}
  \right)\right|=\left|\boldsymbol\Sigma_2\right|\left|\boldsymbol\Sigma_0-\boldsymbol\Sigma_{20}^T\boldsymbol\Sigma_2^{-1}\boldsymbol\Sigma_{20}\right|
\]
The GV-optimal increment is the design $\xi_2$ that minimizes the determinant of the kriging covariance matrix of the second stage $\left|\boldsymbol\Sigma_0^+\right|=\left|\boldsymbol\Sigma_0-\boldsymbol\Sigma_{20}^T\boldsymbol\Sigma_2^{-1}\boldsymbol\Sigma_{20}\right|$, which is exactly the second factor of the determinant factorization above. Obviously for a fixed kriging covariance matrix of the first stage minimizing $\left|\boldsymbol\Sigma_0-\boldsymbol\Sigma_{20}^T\boldsymbol\Sigma_2^{-1}\boldsymbol\Sigma_{20}\right|$ is equivalent to maximizing $\left|\boldsymbol\Sigma_2\right|$ what completes the proof.$\qquad\blacksquare$
\end{myproof}
\medskip

\begin{rem}
Theorem \ref{thm2} is the reason for the great computational benefit of using incremental designs. It reduces the computation of the usual criterion function, which beside some matrix inversions demands the computation of the determinant of a $(m\times m)$-matrix, to the computation of the determinant of a $(l\times l)$-matrix with $l\ll m$. This fact enables an increase of the set of points $X=\{x_1,\ldots,x_{k+l+m}\}$ to an arbitrary size $N=k+l+m$ without raising the computational demands. Actually, the problem of finding $\arg\max_{\xi_2}\left|\boldsymbol\Sigma_2\right|$ is known to be NP-hard, see \cite{Ko_95}, the additional demand of computing the determinant of a $(m\times m)$-matrix would make it intractable already for moderate $m>1000$.
\end{rem}
\medskip

\begin{rem}
Theorem \ref{thm2} can even be used for an efficient computation of GV-optimal designs of a given size $k+l$: We start with some minimal preliminary design, i.e. a design of minimal necessary size $k=p$ where $p$ is the number of deterministic functions of the locations $x\in\xi$: $\textbf{f}(x)=\left(f_1(x)\;f_2(x)\;\cdots\;f_p(x)\right)$ used in the linear predictor. The design points can even be chosen randomly and its kriging covariance matrix is the basis for the computation of the increment $\xi_2$ as above. After this incremental step the design is reduced to size $k_1\ge k$. Also the $k_1$ design points are chosen randomly out of the incremental design of size $k+l$. On the basis of these $k_1$ points we again compute the GV-optimal increment to end up with a design of size $k+l$. As the computational effort is small, these decremental and incremental steps may be repeated many times. If $k$ and $l$ are of moderate size we may even loop systematically through all $k_1$-combinations of the $k+l$ design points.
\end{rem}
\medskip

\begin{rem}
By applying the above incremental step several times we may also construct highly efficient sequential designs which accounts for active learning.
\end{rem}
\smallskip

We can now formulate a similar procedure for efficient computation of increments for V-optimal designs.

\begin{coro}\label{coro3}
The V-optimal increment for the second stage given a design at the first stage is
\begin{equation}\label{updateVcriterion}
  \xi_{\textrm{V}}^{inc}=\arg\max_{\xi_2}\textrm{tr}\left(\boldsymbol\Sigma_2\right)+\textrm{tr}\left(\boldsymbol\Sigma_2^{-1}\boldsymbol\Sigma_{20}\boldsymbol\Sigma_{20}^T\right)
\end{equation}
\end{coro}

\begin{myproof}{Corollary}{\ref{coro3}}
For a given design at the first stage the trace of the corresponding kriging covariance matrix is fixed and
\[\textrm{tr}\left(
    \begin{array}{cc}
      \boldsymbol\Sigma_2 & \boldsymbol\Sigma_{20} \\
      \boldsymbol\Sigma_{20}^T & \boldsymbol\Sigma_0 \\
    \end{array}
  \right)=\textrm{tr}\left(\boldsymbol\Sigma_2\right)+\textrm{tr}\left(\boldsymbol\Sigma_0\right)
\]
The V-optimal increment is the design $\xi_2$ that minimizes the trace of the kriging covariance matrix of the second stage $\textrm{tr}\left(\boldsymbol\Sigma_0^+\right)=\textrm{tr}\left(\boldsymbol\Sigma_0\right)-\textrm{tr}\left(\boldsymbol\Sigma_{20}^T\boldsymbol\Sigma_2^{-1}\boldsymbol\Sigma_{20}\right)$. Obviously for a fixed kriging covariance matrix of the first stage minimizing $\textrm{tr}\left(\boldsymbol\Sigma_0\right)-\textrm{tr}\left(\boldsymbol\Sigma_{20}^T\boldsymbol\Sigma_2^{-1}\boldsymbol\Sigma_{20}\right)$ is equivalent to maximizing $\textrm{tr}\left(\boldsymbol\Sigma_2\right)+\textrm{tr}\left(\boldsymbol\Sigma_{20}^T\boldsymbol\Sigma_2^{-1}\boldsymbol\Sigma_{20}\right)= \textrm{tr}\left(\boldsymbol\Sigma_2\right)+\textrm{tr}\left(\boldsymbol\Sigma_2^{-1}\boldsymbol\Sigma_{20}\boldsymbol\Sigma_{20}^T\right)$ which completes the proof.$\qquad\blacksquare$
\end{myproof}
\medskip

\begin{rem}
Corollary \ref{coro3} is the reason for the computational benefit of using incremental designs. It reduces the computation of the usual criterion function which demands the matrix inversions of one $(k+l)\times(k+l)$- and one $(p\times p)$-matrix and the computation of the new kriging covariance matrix (10 matrix multiplications of which 5 involve matrices with $m$ as one dimension) to the computation of the inverse of a $(l\times l)$-matrix and 2 matrix multiplications of which only 1 involves a $(l\times m)$-matrix. As $l\ll m$ this reduces the computational effort to roughly one third.

Just as Theorem \ref{thm2}, Corollary \ref{coro3} can be used for an efficient computation of designs close to V-optimality of a given size $k+l$. The computational benefit of Corollary \ref{coro3} cannot be compared to the improvement of Theorem \ref{thm2}, as we have to limit the number of incremental and decremental steps here and we thus have no guarantee to end up with the V-optimal design.
\end{rem}

\subsection{GV-optimal designs in design spaces dense in $\mathbb{R}^d$}\label{Ss:DENSE}
In subsection (\ref{S:BLUP}) we state that the allowable choice of the designs $\xi$ is from a fixed finite set of $N=k+m$ points $X=\left\{x_1,\ldots,x_{k+m}\right\}$. Usually $k\ll m$ and going into details of Corollary 1, Remark 2 we may even extend the number of non-design-points $m$ to any (integer) size.

The reason for this remarkable possibility is that for the computation of the GV-optimal increment $\xi_2$ of a given design $\xi_1$ we do not even need the kriging covariance matrix of the first stage which would be of enormous dimension $(l+m)\times(l+m)$, we just need the $(l\times l)$-block $\boldsymbol\Sigma_2$ which is
\[\boldsymbol\Sigma_2=\mathbf{W}_{12}\mathbf{C}_1\mathbf{W}_{12}^T-\mathbf{W}_{12}\mathbf{C}_{12}-\mathbf{C}_{12}^T\mathbf{W}_{12}^T+\mathbf{C}_2
\]
The dimensions of all of the above matrices are just $k$ or $l$. $\mathbf{W}_{12}$ are the weights of $\mathbf{Y}_1$ for the prediction of $\mathbf{Y}_2$ in the first stage:
\[\mathbf{W}_{12}=\mathbf{F}_2\mathbf{B}+\mathbf{C}_{12}^T\mathbf{A}
\]
with $\mathbf{B}=\left(\mathbf{F}_{1}^T\mathbf{C}_{1}^{-1}\mathbf{F}_{1}\right)^{-1} \mathbf{F}_{1}^T\mathbf{C}_{1}^{-1}$ and $\mathbf{A}=\mathbf{C}_{1}^{-1}\left(\mathbf{I}_k-\mathbf{F}_{1}\mathbf{B}\right)$. The dimensions of all these matrices are only $k$, $l$ and $p$, i.e. for the computation of $\boldsymbol\Sigma_2$ we just need small matrices independent of the number $m$.

The GV-criterion function is the determinant of the kriging covariance matrix, if in the first stage this determinant is $D$, then the criterion function of the incremental design is $\frac{D}{|\boldsymbol\Sigma_2|}$.

In the decremental step we remove $l_1=k+l-k_1$ design points from the incremental design, the according $(l_1\times l_1)$-block of the kriging covariance matrix of the first stage is $\boldsymbol\Sigma_2^*$ and may be computed as shown above.

The determinant of the kriging covariance matrix after the decremental step then is $\frac{D}{|\boldsymbol\Sigma_2|}\cdot|\boldsymbol\Sigma_2^*|$. So, not knowing the value of the GV-criterion during the search for the GV-optimal design is not crucial as it suffices to know the $\boldsymbol\Sigma_2$-blocks according to the increments and decrements to minimize the GV-criterion. Thus in principle we can make the grid as dense as desired, as long as the number of points is finite.
%Of course this situation is not completely satisfactory because the design space should not be limited to a finite number of points. But
There is reasonable hope that the described method can be generalized to continuous design spaces, which we defer to future research.

%_________________________________________________________________________________________________________________
\section{Efficiency of GV-optimal designs}\label{S:EFF}
As mentioned above it turns out that GV-optimal designs are highly efficient with other design criteria which will be discussed in subsection (\ref{Effother}).

Additionally Corollary 1 allows a simple, fast and computationally very efficient calculation of incrementally constructed designs that are close to be GV-optimal. The efficiency of these incremental designs will be discussed in subsection (\ref{Effincr}).

%---------------------------------------------------------------------------------------------------
\subsection{Efficiency with respect to other design criteria}\label{Effother}
Traditional criteria for optimal designs for prediction are usually concerned with the variance of predictions, i.e. we could also title this subsection with "Efficiency with respect to variance-based criteria". Here we compare our GV-optimal designs with G- and V-optimal designs with the help of the \textbf{relative efficiency}, a very common concept in comparing designs, see eg. \cite{lopez-fidalgo_optimal_2023}, p.17.

Let $\Phi_G$, $\Phi_V$ and $\Phi_{GV}$ be the criterion function for G- , V- and GV-optimality respectively and $\xi_G$, $\xi_V$ and $\xi_{GV}$ be the G-, V- and GV-optimal designs of the same size for the prediction of the same number of points. The GV-optimal design $\xi_{GV}$ minimizes $\Phi_{GV}$, as the other optimal designs minimize their corresponding design criteria. Then e.g. the relative G-efficiency of the V-optimal design is
\[E_G\left(\xi_G,\xi_V\right)=\frac{\Phi_G\left(\xi_G\right)}{\Phi_G\left(\xi_V\right)}.
\]
We always have $E_{\;\cdot}\left(\cdot,\cdot\right)<=1$ and the relative efficiency of a design $\xi$ gives the factor the criterion function of $\xi$ may be decreased if we switch to the optimal design.
These relative efficiencies are scale invariant though the effect of scaling is different for the GV-criterion function on the one and the G- and V-criterion functions on the other hand. The kriging covariance matrix of scaled responses $s\cdot\mathbf{Y}$ is $s^2\boldsymbol\Sigma$ which affects G- and V-criterion functions the same: $\max \textrm{diag}\left(s^2\boldsymbol\Sigma\right)=s^2\max \textrm{diag}\left(\boldsymbol\Sigma\right)$ and $\textrm{tr}\left(s^2\boldsymbol\Sigma\right)=s^2\textrm{tr}\left(\boldsymbol\Sigma\right)$. The GV-criterion can be made insensitive to scaling by applying $\sqrt[k]{|\boldsymbol\Sigma|}$ instead without changing the designs.

The relative efficiencies are not affected by this scaling. \textsl{I.e.}, if $\boldsymbol\Sigma_{GV}$, $\boldsymbol\Sigma_{G}$ and $\boldsymbol\Sigma_{V}$ are the kriging covariance matrices for the originally unscaled data of the GV-, G- and V-optimal designs, then the GV-efficiency of the G- and V-optimal designs respectively for scaled data are
\[E_{GV}\left(\xi_{GV},\xi_G\right)=\frac{s^m\sqrt{|\boldsymbol\Sigma_{GV}|}}{s^m\sqrt{|\boldsymbol\Sigma_G|}}\qquad\qquad E_{GV}\left(\xi_{GV},\xi_V\right)=\frac{s^m\sqrt{|\boldsymbol\Sigma_{GV}|}}{s^m\sqrt{|\boldsymbol\Sigma_V|}}\quad,
\]
i.e. arbitrary scaling does not change the relative GV-efficiencies. The same is true also for relative G- and V-efficiency.

%_________________________________________________________________________________________________________________
\section{A representative example}\label{S:EX}

Let us demonstrate the typical relative efficiencies on the basis of the following settings, computations of many other differently adjusted models and designs yield similar results.

The design space was chosen 2-dimensional on a regular grid, $\SX=\{1,2,\ldots,17\}^2$, optimal designs were computed for the Matern covariance model with all combinations of range parameters $\phi\in\{0.1, 0.5, 0.75, 1, 1.5, 2, 3, 4, 5\}$ and smoothness parameter $\kappa\in\{0.25, 0.5,1,1.5,2,2.5\}$. The variance as scaling parameter $s$ was chosen such that for each of the 54 combinations of $\phi$ and $\kappa$ the design criterion $|\boldsymbol\Sigma|$ of the GV-optimal 12-point design is one. Designs of size $6$ (with quadratic trend $7$), $9$ and $12$ were computed for linear and quadratic trend functions.

As can be seen in tables \ref{efflin} and \ref{effquad}, GV-optimal designs are reasonably efficient with respect to the G- and V-criterion. Here for every covariance parameter combination the GV-, G- and V-optimal designs were determined and then for each optimal design the relative efficiencies with respect to the other design criteria were computed. This was here done for a linear trend function and for designs of size 6, 9 and 12 respectively. Finally, the relative efficiencies of the optimal designs were averaged over all 54 covariance parameter combinations. The lines of the tables correspond to GV-, G- and V-optimal designs, and e.g. the mean relative efficiency of 9-point GV-optimal designs with respect to the G-criterion is 0.9707 which means that the maximum kriging variance of G-optimal designs is on average only 97\% of the maximum kriging variance of GV-optimal designs.

\begin{table}[htb]
  \centering
  \begin{tabular}{|r|ccc||ccc||ccc|}
  \hline
  linear & \multicolumn{3}{c||}{6 points} & \multicolumn{3}{c||}{9 points} & \multicolumn{3}{c|}{12 points} \\
  % after \\: \hline or \cline{col1-col2} \cline{col3-col4} ...
  \cline{2-10}
  trend & $E_{GV}$ & $E_G$ & $E_V$ & $E_{GV}$ & $E_G$ & $E_V$ & $E_{GV}$ & $E_G$ & $E_V$ \\
   \hline
  $\xi_{GV}$ & 1 & 0.9050 & 0.9151 & 1 & 0.9707 & 0.9501 & 1 & 0.9385 & 0.9181  \\
  $\xi_{G}$ & 0.9328 & 1 & 0.9704 & 0.9315 & 1 & 0.9781 & 0.9639 & 1 & 0.9606  \\
  $\xi_{V}$ & 0.9010 & 0.9025 & 1 & 0.8756 & 0.9370 & 1 & 0.9316 & 0.9362 & 1 \\
  \hline
\end{tabular}
  \caption{\footnotesize Average relative efficiencies for linear trend over 54 different parameter combinations of Matern covariance models.} \label{efflin}
\end{table}
\bigskip

\begin{table}[htb]
  \centering
  \begin{tabular}{|r|ccc||ccc||ccc|}
  \hline
  quadratic & \multicolumn{3}{c||}{7 points} & \multicolumn{3}{c||}{9 points} & \multicolumn{3}{c|}{12 points} \\
  % after \\: \hline or \cline{col1-col2} \cline{col3-col4} ...
  \cline{2-10}
  trend & $E_{GV}$ & $E_G$ & $E_V$ & $E_{GV}$ & $E_G$ & $E_V$ & $E_{GV}$ & $E_G$ & $E_V$ \\
  \hline
  $\xi_{GV}$ & 1 & 0.7951 & 0.8997 & 1 & 0.9928 & 0.9606 & 1 & 0.9100 & 0.9059 \\
  $\xi_{G}$ & 0.9173 & 1 & 0.9330 & 0.9712 & 1 & 0.9702 & 0.9360 & 1 & 0.9614 \\
  $\xi_{V}$ & 0.8175 & 0.7303 & 1 & 0.8739 & 0.8875 & 1 & 0.8966 & 0.9038 & 1 \\
  \hline
  \end{tabular}
  \caption{\footnotesize Average relative efficiencies for quadratic trend over 54 different parameter combinations of Matern covariance models.} \label{effquad}
\end{table}

\subsection{Efficiency of incrementally assembled designs}\label{Effincr}

As already mentioned incrementally assembled designs are very efficient with respect to the GV-criterion. In the following discussion we will always start with some $k$-point design $\xi_k$ adding a single increment of size $l$. The result will only be the GV-optimal design of size $(k+l)$ if we start in the first stage with a $k$-point design $\xi_k\subset\xi_{k+l}$ which is very improbable if we do not utilize additional knowledge. Usually we also will not end up in the GV-optimal $(k+l)$-point design if we start with the GV-optimal $k$-point design, but the result will be very close to GV-optimality. How close the incrementally constructed design is to GV-optimality depends on the choice of the $k$-point starting design. Of course we may take advantage of prior knowledge about properties of GV-optimal designs, e.g. if the design region is the unit square as in our simulation examples, we know that the GV-optimal design for a linear or quadratic trend will always have design points in the corners and the edges of the design region. Choosing such plausible points for the $k$-point starting design will almost always yield GV-optimal $(k+l)$-point designs that are constructed with a single incremental step.

Here we followed two ideas:
\begin{itemize}
  \item start with a GV-optimal design of (small) size $k$;
  \item start with a plausible design of (small) size $k$, i.e. with design points that most likely are elements of GV-optimal designs of arbitrary size.
\end{itemize}
It turns out that both ideas yield very efficient designs especially if the increment (the number of additional design points) is not too small.

To exemplify this efficiency we again used Matern covariance models with 54 different parameter combinations and a quadratic trend.
In the first simulation series we started with a plausible 6-point design, i.e. 4 design points at the corners and 2 points on opposite margins of the unit square. Then we added the optimal increment of 6 design points as described above. The mean GV efficiency of these incremental designs was 0.9911, the median efficiency was even 100\%.

In the second series of simulations we started with the 7-point GV-optimal designs for each parameter combination respectively and added the optimal 5-point increment. Here the mean GV efficiency was 0.9862 and the median efficiency again 100\% indicating a satisfactory performance of this simple incremental method.

We applied the same concept to linear trends as well. The 4 corners of the design region were chosen as plausible starting design, and the increment of size 8 always yielded the GV-optimal 12-point design. Starting with 6-point GV-optimal designs for each parameter combination respectively and adding the optimal 6-point increment resulted in a mean GV-efficiency of 0.9881, again the median efficiency was 100\%.

Note that the above mean and median efficiencies are just for a design built with a single incremental step. Of course we may always append a few decremental-incremental steps to guaranty GV-optimality. This approach is analyzed in the next section.
\begin{figure}
\begin{center}
\includegraphics[width=\textwidth]{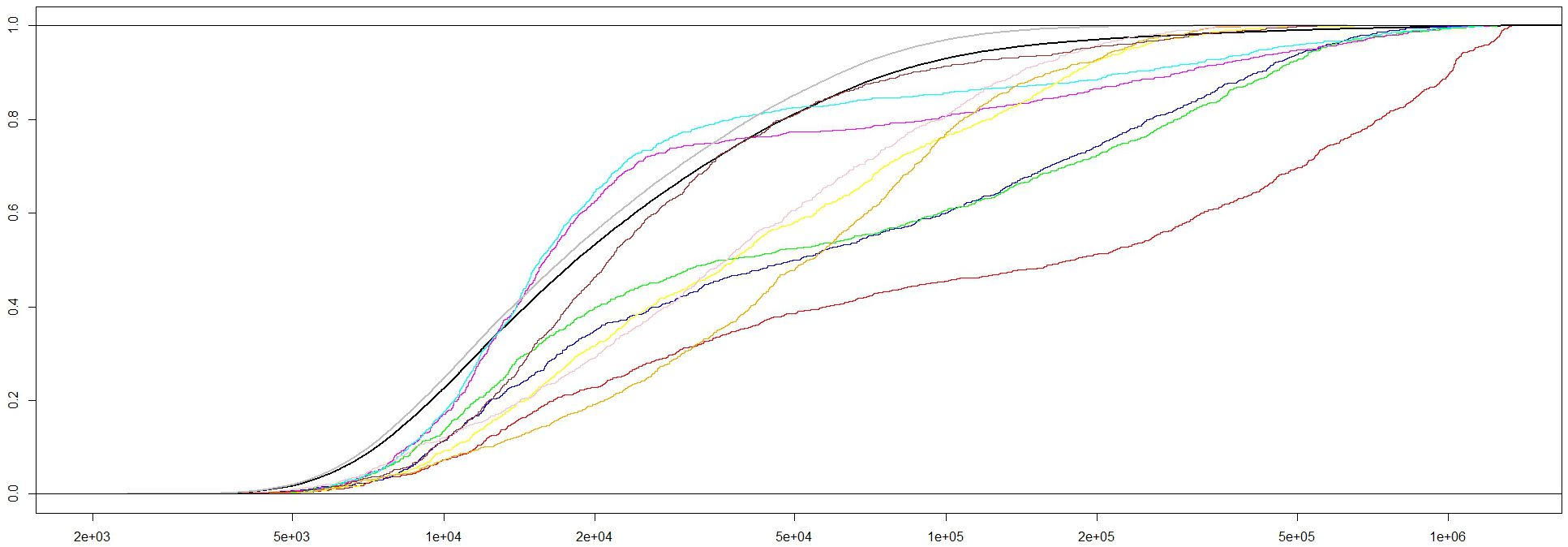}
\end{center}
\caption{\footnotesize Distribution of number of computations of the criterion function till convergence to the GV-optimal design including (black line) and excluding (grey line) distributions for ``adverse'' parameter combinations (colored lines) - mind the log-scale.}
\label{F:distrcalls}
\end{figure}

\subsection{Efficiency of incrementally-decrementally assembled designs}
The above described method of systematically discarding design-points after each incremental step has an efficiency of 100\%. For each of the 54 combinations of the $\kappa$ and $\phi$ parameters of the Matern covariance model (see above) we started 1000 times with a random design and ended with the GV-optimal design every time just with 26 exceptions (where a design with efficiency 99.9\% was found instead of the optimal design). It turned out that theses exceptions were all for designs with three special parameter combinations of $\kappa$ and $\phi$ which seem to be adverse for finding the GV-optimal design on the chosen grid. For these three parameter combinations also the average number of required computations of the criterion function was incomparably higher then for other parameter values (see Table \ref{T:callsGV}). The reason was obviously that the design points were limited to the unfavourable $(17\times17)$ grid. Changing to a $(33\times33)$ grid  solved this problem. With the finer grid we started 200 times for each of the 54 parameter combinations of $\kappa$ and $\phi$ with a random design and found the optimal designs without exception. Also the number of computations of the criterion function till convergence was distributed more uniform than with the $(17\times17)$ grid. Though there were some parameter combinations which needed clearly more function calls, for all these cases the second best design always was very efficient and the search algorithm now and then got stuck at these designs before finding the optimum.

\subsubsection{Speed of convergence to GV-, G- and V-optimal designs}\label{Speed}

The speed of convergence was measured in absolute time and in the number of required computations of the criterion function.

In Table \ref{T:callsGV} we can see the median number of computations of the criterion function (in this case the determinant of the $(6\times6$)-matrix $\boldsymbol\Sigma_2$) needed to find the GV-optimal design. The overall median number of calls of the criterion function was 17222, the computation time for finding 54.000 times the GV-optimal designs was 50.14 hours; i.e. 3.34 seconds per optimal design.

The distributions of the number of criterion function evaluations turned out to be positively skewed, the cdf of such distributions for selected parameter combinations of $\kappa$ and $\phi$ is depicted in Figure \ref{F:distrcalls}. There were 3 parameter combinations where the corresponding distributions of function calls were striking.  Discarding these extreme distributions reduced the average computation time for one GV-optimal design to 2.18 sec. Changing to a grid twice as fine as the original $(17\times17)$ points to which the design was limited solved the problem (see above). The overall median number of calls of the criterion function increased to 24877 with the $(33\times33)$ grid which was caused by the halved step size in the neighbourhood search at the finer grid. Here continuous optimization algorithms with variable step size promise an improvement (see Section \ref{Ss:DENSE}).
\begin{table}
  \centering
{\footnotesize
\begin{tabular}{|l||rrrrrrrrr|}
  \hline
  % after \\: \hline or \cline{col1-col2} \cline{col3-col4} ...
  $\kappa$ & $\phi=0.1$ & $\phi=0.5$ & $\phi=0.75$ & $\phi=1$ & $\phi=1.5$ & $\phi=2$ & $\phi=3$ & $\phi=4$ & $\phi=5$ \\
  \hline
  \hline
  0.25 & 8614.5 & 36704.0 & 15789.0 & 16425.0 & 18160.0 & 20516.0 & 22970.5 & 25423.5 & 27438.0 \\
  0.5 &  8821.0 & 29894.0 & 12355.0 & 16146.0 & 20333.0 & 24225.0 & 36111.5 & 53231.5 & 21380.0 \\
  1 &    8895.0 & 25630.0 & 11937.5 & 15876.0 & 21115.0 & 29033.5 & 33065.0 & 36066.0 & 14808.5 \\
  1.5 &  9052.0 & 27106.0 & 13237.5 & 17169.5 & 19459.5 & 33542.0 & 21889.0 & 14687.0 & 15538.0 \\
  2 &    8943.0 & 27852.0 & 15681.0 & 15581.5 & 20344.0 & 34206.0 & 21146.5 & 14710.5 & 179657.0 \\
  2.5 &  8702.0 & 27995.5 & 22359.0 & 17660.0 & 19273.0 & 33752.5 & 50351.5 & 15931.0 & 37856.5 \\
  \hline
\end{tabular}
  \caption{\footnotesize Median number of computations of the criterion function until convergence to the GV-optimal design. }\label{T:callsGV}
}
\end{table}

The GV-optimal designs are found incomparably faster than corresponding G- and V-optimal designs even if the allowable choice of designs is from a moderate number of points.

Searching for G-optimal designs could only be tried 10 times for each of the 54 combinations of the $\kappa$ and $\phi$ parameters and 368 of these 540 tries failed, because we had to stop the search algorithm (a combination of neighboring point exchanges and simulated annealing, same algorithm was used to find the GV-optimal designs) after 500 iterations because of the huge expenditure of time. The reason for that was partly the much larger computational effort but also a much slower convergence to the optimum, i.e. the criterion function ($\max\textrm{diag}\left(\boldsymbol\Sigma_0-\boldsymbol\Sigma_{20}^T\boldsymbol\Sigma_2^{-1}\boldsymbol\Sigma_{20}\right)$) had to be called much more frequently than in the search for GV-optimal designs. The mean G-efficiencies of the 540 found designs was 0.991, for $\kappa$ and $\phi$ parameter combinations corresponding to high correlated data the mean G-efficiencies of the found designs were considerably smaller (with a minimum of 0.927 for $\kappa=2.5$ and $\phi=5$).

In Table \ref{T:callsG} the median number of calls of the criterion function for each of the 54 combinations of the $\kappa$ and $\phi$ parameters is reported. The overall median number of calls of the criterion function was 5.101.156 (296.2 times as often as for GV-optimal designs), the computation time for finding 540 times the G-optimal designs was 761.5 hours, i.e. 1.41 hours per optimal design which is 1520 times as long as for one GV-optimal design.

\begin{table}
  \centering
{\footnotesize
\begin{tabular}{|l||rrrrrrrrr|}
  \hline
  $\kappa$ & $\phi=0.1$ & $\phi=0.5$ & $\phi=0.75$ & $\phi=1$ & $\phi=1.5$ & $\phi=2$ & $\phi=3$ & $\phi=4$ & $\phi=5$ \\
  \hline
  \hline
  0.25 & 5257485 & 5656141 & 5054189 & 1283471 & 5187172 & 5251948 & 5164463 & 4815720 & 5856377 \\
  0.5 &  5371069 & 6300232 & 4827949 & 3486381 & 3627394 & 4942996 & 5708568 & 5169052 & 1830513 \\
  1 &    4293287 & 5535166 & 5426545 & 4163447 & 5408354 & 5452206 & 3041665 & 5309562 & 7470061 \\
  1.5 &  4465675 & 4167031 & 5375233 & 4891069 & 5037755 & 5178237 & 5707364 & 4545278 & 6490938 \\
  2 &    5457978 & 5412340 & 4255818 & 2904401 & 5235663 & 5193218 & 6097624 & 6819590 & 6976215 \\
  2.5 &  4748402 & 5251379 & 2365235 & 5010149 & 6236549 & 3721464 & 6237881 & 5627650 & 6549851 \\
  \hline
\end{tabular}
  \caption{\footnotesize Median number of computations of the criterion function until convergence to the G-optimal design.}\label{T:callsG}
}
\end{table}

V-optimal designs are somehow found easier than G-optimal designs (because we may apply Corollary 2). We managed to search the V-optimal design 250 times for each of the 54 combinations of the $\kappa$ and $\phi$ parameters. In table \ref{T:callsV} we can see the median number of computations of the criterion function (in this case $\textrm{tr}\left(\boldsymbol\Sigma_2\right)+\textrm{tr}\left(\boldsymbol\Sigma_2^{-1}\boldsymbol\Sigma_{20}\boldsymbol\Sigma_{20}^T\right)$) needed to find the V-optimal design. Also here the average number of calls of the criterion function was clearly larger than for the GV-optimal design. The overall median number of calls of the criterion function was 108928.5 (6.3 times as often as for GV-optimal designs), the computation time for finding 13.500 times the V-optimal designs was 240.7 hours, i.e. 64.2 seconds per optimal design which is 20 times as long as for one GV-optimal design.

\begin{figure}
\begin{center}
\includegraphics[width=\textwidth]{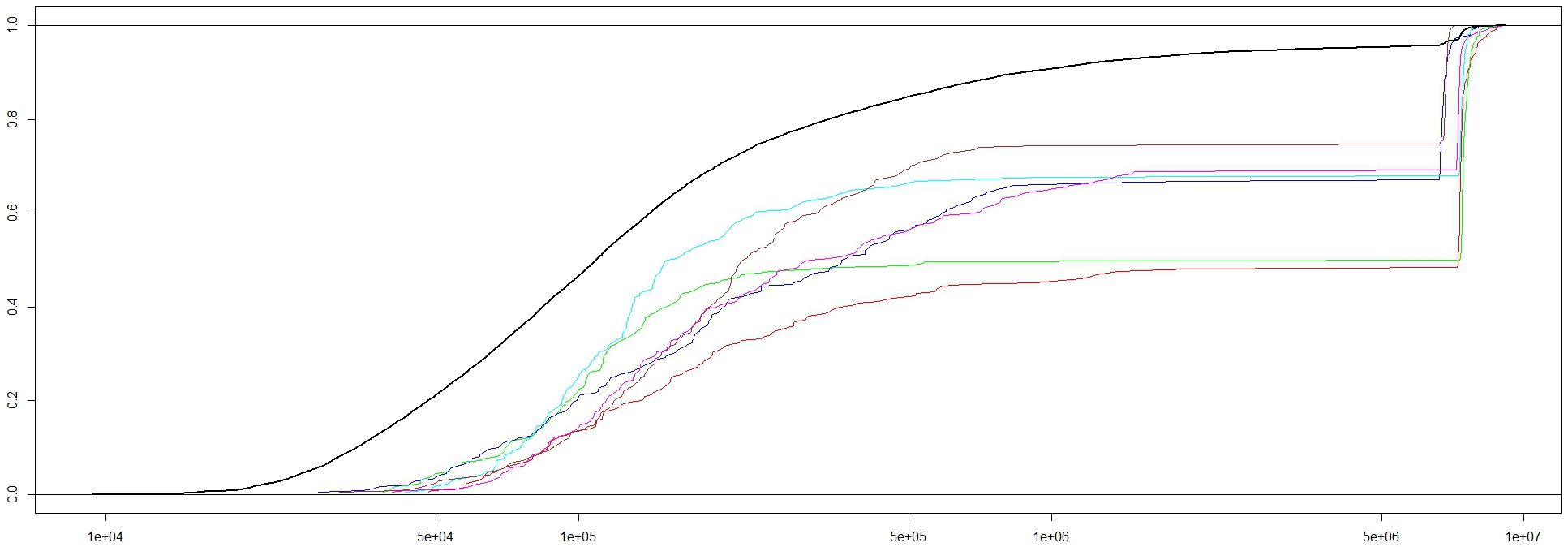}
\end{center}
\caption{\footnotesize Empirical cdf of the number of computations of the criterion function till convergence to the V-optimal design (black line) and for parameter combinations where the optimal design was not always found within a maximal number of tries (colored lines) - mind the log-scale.}
\label{F:distrcallsV}
\end{figure}

Of the 13.500 tries to find the V-optimal design 577 failed, that is 4.3\% (almost exactly 100 times more than for GV-optimal designs). This has also an impact on the cdf of the positively skewed distribution of the number of calls of the criterion functions until the optimal designs were found (Figure \ref{F:distrcallsV}). As with G-optimal designs we stopped the search algorithm after 500 iterations, the cdf's for parameter combinations where the optimal designs were not found within this maximal number of iterations are depicted as colored lines.

\begin{table}
  \centering
{\footnotesize
\begin{tabular}{|l||rrrrrrrrr|}
  \hline
  $\kappa$ & $\phi=0.1$ & $\phi=0.5$ & $\phi=0.75$ & $\phi=1$ & $\phi=1.5$ & $\phi=2$ & $\phi=3$ & $\phi=4$ & $\phi=5$ \\
  \hline
  \hline
  0.25 & 37326.0 & 47700.5 & 71195.0 & 73747.5 & 148730.5 & 343129.0 & 157580.5 & 312331.5 & 224414.0 \\
  0.5 &  32191.5 & 42299.5 & 59498.5 & 90473.0 & 1662530.5 & 386891.0 & 104161.5 & 101856.5 & 98935.5 \\
  1 &    31478.5 & 49869.0 & 61041.5 & 415871.5 & 200776.5 & 154566.5 & 119494.5 & 124363.0 & 120898.0 \\
  1.5 &  30625.0 & 48430.5 & 84230.0 & 82082.5 & 186643.5 & 297843.0 & 265400.5 & 143954.0 & 143626.5 \\
  2 &    30864.0 & 62543.0 & 69631.0 & 144874.0 & 364671.0 & 187267.5 & 257116.5 & 325020.0 & 157888.0 \\
  2.5 &  30746.5 & 78648.5 & 87787.0 & 102481.5 & 259893.0 & 137820.5 & 276008.5 & 7259052.5 & 7333013.0 \\
  \hline
\end{tabular}
  \caption{\footnotesize Median number of computations of the criterion function until convergence to the V-optimal design.}\label{T:callsV}
}
\end{table}

%_________________________________________________________________________________________________________________
\section{Real illustrative example: temperature prediction in Upper Austrian municipalities}\label{S:RWEX}

The province of Upper Austria is partitioned in 438 rural and urban municipalities with considerable topographical differences ranging from lowlands in the center and hill country in the north to high-altitude mountains in the south. As temperature and its spatial variation is strongly influenced by the topography the simultaneous prediction of temperatures at all principle locations of the 438 municipalities is challenging.

Currently there exist 36 meteorological stations in Upper Austria that may be taken as data source for temperature prediction in the 438 municipalities. A natural question is whether the current network can be improved by relocation of the station and/or we can even reduce the size of the network without loss of accuracy.

We model the expected monthly mean temperatures $t_{ij}$ with the elevation of the measurement location $el_i$ as external drift which is in line with \cite{hudson_mapping_1994}:
\[\mu(t_{ij})=\beta_j+\beta_i\cdot el_i,
\]
where $j$ indicates the month and $i$ the location of the measurements. We further use an anisotropic Matérn covariance model to describe the spatial interdependencies of temperatures measured in the same time period. The covariance parameters have been estimated with the \verb"likfit" function of the R package \verb"GeoR"(\cite{geoR_2001}).

The learning data for parameter estimation were the daily mean temperatures of all meteorological stations in Upper Austria in the period from 2000-01-01 until 2023-10-25. The data are publicly available at the \cite{GeoSphere2023}. The coordinates and elevations of the 438 municipalities are also publicly available at the DORIS webOffice (\url{https://www.doris.at/})

The parameter estimates confirm the environmental temperature lapse rate of $\sim 6.5$°C/km (\cite{ICAO1993}, \cite{thompson1998}) and are similar for all months except the winter period when the phenomenon of temperature inversion (\cite{NOAA2023}) may be observed frequently.

Here the showcase results for the month June are presented, other months are comparable. The design region is the set of 438 locations corresponding to the principle localities of the 438 Upper Austrian administrative municipalities, actually 36 of these locations are the base of meteorological stations (\url{https://bitly.ws/ZFpC}). We want to evaluate the prediction quality of this actual meteorological network by means of the GV-criterion and compare it with a virtual network positioned at the locations of the GV-optimal design of the same or a reduced size. I.e., initially we have $k=36$ and $m=438-36=402$ locations for simultaneous prediction.

\begin{figure}
\begin{center}
\includegraphics[width=.8\textwidth]{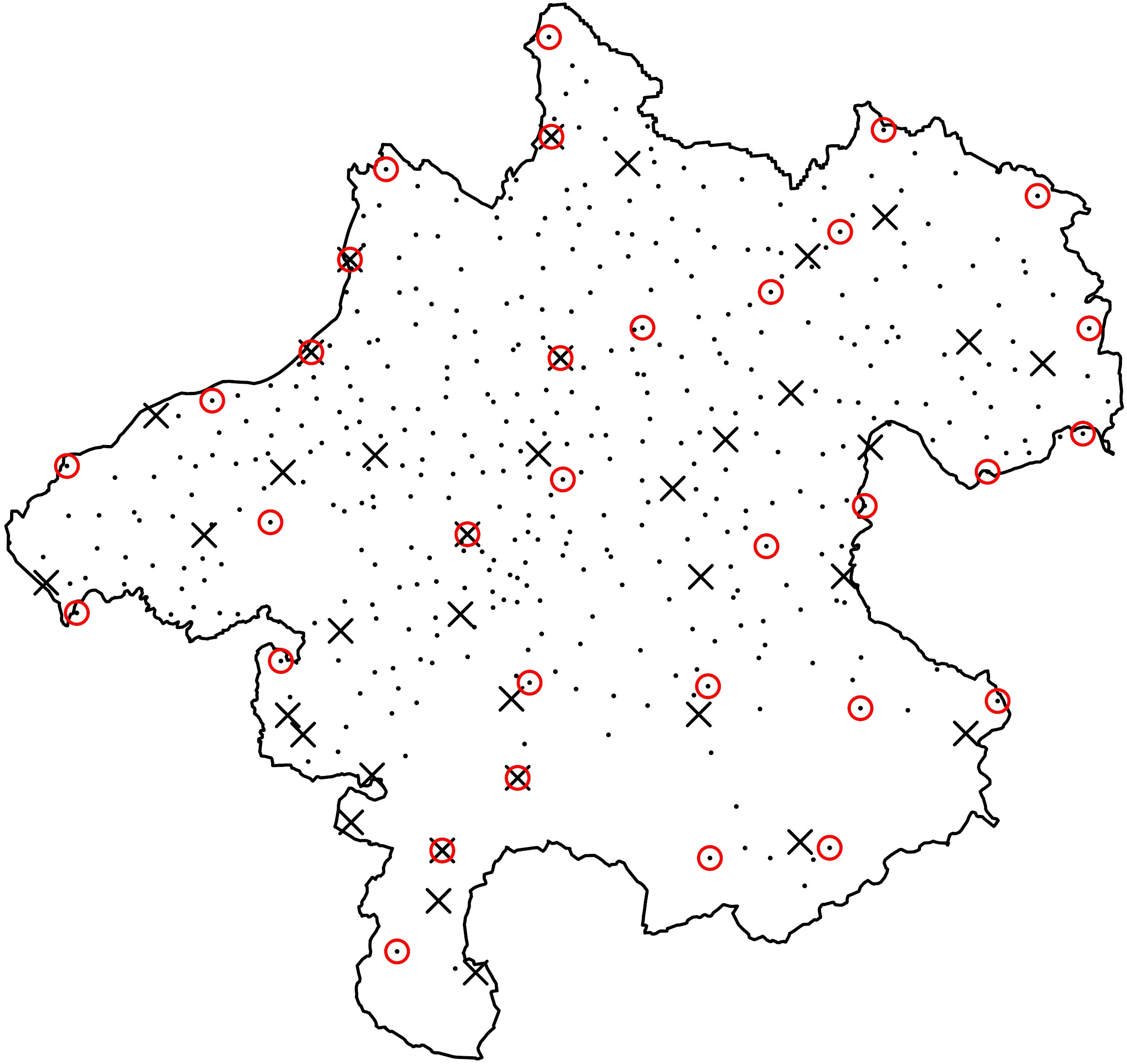}
\end{center}
\caption{\footnotesize Map of the 438 municipalities (dots), the current 36 meteorological stations (crosses), the locations of the 32-point GV-optimal design (red circles) for a meteorological network for simultaneous prediction of temperatures in the 438 municipalities of Upper Austria.}
\label{F:stations}
\end{figure}

We then further reduced the size step by step down to $k=32$ until which we yielded the same accuracy according to the GV-criterion as from the the initial network. This resulting GV-optimal design as well as the actual meteorological network and all 438 locations are displayed in Figure \ref{F:stations}. Remarkably 7 locations of the optimal design are already base of a meteorological station and other 7 locations of the optimal design are within a range of 5km from an actual station although to our knowledge no statistical analysis was involved in determining the current network.
While the reduction of just 4 stations from 36 down to 32 may seem as a disappointment one must not forget that the number of prediction locations had to increase from 402 to 406 making the task more difficult.

%_________________________________________________________________________________________________________________
\section{Conclusions}\label{S:CONCL}

With $GV$-optimality we have introduced a novel design criterion for simultaneous kriging prediction, which considers the whole prediction covariance matrix. As was shown by the real-world example there are indeed practical problems requiring simultaneous rather than individual prediction. In such situations, the presented new criterion is a natural answer and more adequate and useful.

In terms of robustness, it has been demonstrated that GV-optimal designs exhibit considerable efficiency compared to designs optimized based on other criteria. The criterion function is notably smoother compared to other criteria, meaning that slight modifications to the design do not lead to significant alterations in the criterion function. Interestingly, this might also be the main reason that GV-optimal designs are found much faster than designs optimal with respect to other criteria (which is subject to actual and future research).

%The predictions based on GV-optimal designs are not computed different from other designs. In this sense predictions will not differ from other designs. The difference lies in the expected accuracy and the expected overall variation of the prediction errors. This smaller overall range of the prediction errors is the main benefit of GV-optimal designs.

Furthermore we have shown that efficient incremental construction methods are available, which makes the criterion particularly attractive for big data and higher dimensional contexts. For instance it lends itself naturally combinable with local kriging techniques such as \cite{gramacy_speeding_2016}. These and other extensions will be subject of future research.

\section{Acknowledgements}
This work was partly supported by project INDEX(INcremental Design of EXperiments) I3903-N32 of the Austrian Science Fund(FWF). The third author received support for this research from a fellowship provided by the Spanish Ministry of Universities (PRX22/00578).

We give credits to the Digitales Oberösterreichisches Raum-Informations-System [DORIS] for permitting free use of its public data under the license \url{https://bitly.ws/ZEuP} is given. The file for coordinates and elevations was downloaded from  \url{https://bitly.ws/ZEPI} 2023-10-12.

We are gratefully to two referees and an AE, whose comments lead to a considerable improvement of the paper.

%% The Appendices part is started with the command \appendix;
%% appendix sections are then done as normal sections
 \appendix

%_________________________________________________________________________________________________________________
\section{Proofs}
\subsection{Proof of the formula for the kriging weights of the second stage $\mathbf{W}_2$}
\textbf{Proof for simple kriging:} We first compute kriging weights and the kriging covariance matrix for the first stage and apply the update formulae (\ref{updateW22}) and then compare the results to the kriging weights directly computed for the second stage which are the same what completes the proof.

The kriging weights for the first stage are $\mathbf{W}_{12}=\mathbf{C}_{12}^T\mathbf{C}_1^{-1}$ and $\mathbf{W}_{10}=\mathbf{C}_{10}^T\mathbf{C}_1^{-1}$. Plugging this weights into (\ref{krigingcov}) gives the kriging covariance matrix for the first stage:
\begin{equation}\label{krigcovstage1}
  \left(
    \begin{array}{cc}
      \boldsymbol\Sigma_2 & \boldsymbol\Sigma_{20} \\
      \boldsymbol\Sigma_{20}^T & \boldsymbol\Sigma_0 \\
    \end{array}
  \right)=\left(
    \begin{array}{cc}
      \mathbf{C}_2-\mathbf{C}_{12}^T\mathbf{C}_1^{-1}\mathbf{C}_{12} & \mathbf{C}_{20}-\mathbf{C}_{12}^T\mathbf{C}_1^{-1}\mathbf{C}_{10} \\
      \mathbf{C}_{20}^T-\mathbf{C}_{10}^T\mathbf{C}_1^{-1}\mathbf{C}_{12} & \mathbf{C}_0-\mathbf{C}_{10}^T\mathbf{C}_1^{-1}\mathbf{C}_{10} \\
    \end{array}
  \right)
\end{equation}

This is plugged into the update formula (\ref{updateW22}) to get \\ $\mathbf{W}_{2}=\left(\mathbf{C}_{20}^T-\mathbf{C}_{10}^T\mathbf{C}_1^{-1}\mathbf{C}_{12}\right)\left(\mathbf{C}_2-\mathbf{C}_{12}^T\mathbf{C}_1^{-1}\mathbf{C}_{12}\right)^{-1}$
and\\
$\mathbf{W}_{1}=\mathbf{C}_{10}^T\left(\mathbf{C}_1^{-1}+\mathbf{C}_1^{-1}\mathbf{C}_{12} \left(\mathbf{C}_2-\mathbf{C}_{12}^T\mathbf{C}_1^{-1}\mathbf{C}_{12}\right)^{-1}\mathbf{C}_{12}^T\mathbf{C}_1^{-1}\right)- \mathbf{C}_{20}^T\left(\mathbf{C}_2-\mathbf{C}_{12}^T\mathbf{C}_1^{-1}\mathbf{C}_{12}\right)^{-1}\mathbf{C}_{12}^T\mathbf{C}_1^{-1}$ which should be the kriging weights for the second stage.

Now we compute the weights for the second stage directly:
\[\mathbf{W}=\left(
               \begin{array}{cc}
                 \mathbf{C}_{10}^T & \mathbf{C}_{20}^T \\
               \end{array}
             \right)\left(
                      \begin{array}{cc}
                        \mathbf{C}_1 & \mathbf{C}_{12} \\
                        \mathbf{C}_{12}^T & \mathbf{C}_2 \\
                      \end{array}
                    \right)^{-1}
\]
Using the identity
{\footnotesize
\begin{eqnarray}\label{blockinv}
 \nonumber % Remove numbering (before each equation)
  & & \left(
                      \begin{array}{cc}
                        \mathbf{C}_1 & \mathbf{C}_{12} \\
                        \mathbf{C}_{12}^T & \mathbf{C}_2 \\
                      \end{array}
                    \right)^{-1}= \\
  &=& \left(
                                   \begin{array}{cc}
                                     \mathbf{C}_1^{-1}+\mathbf{C}_1^{-1}\mathbf{C}_{12} \left(\mathbf{C}_2-\mathbf{C}_{12}^T\mathbf{C}_1^{-1}\mathbf{C}_{12}\right)^{-1}\mathbf{C}_{12}^T\mathbf{C}_1^{-1} & -\mathbf{C}_1^{-1}\mathbf{C}_{12}\left(\mathbf{C}_2-\mathbf{C}_{12}^T\mathbf{C}_1^{-1}\mathbf{C}_{12}\right)^{-1} \\
                                     -\left(\mathbf{C}_2-\mathbf{C}_{12}^T\mathbf{C}_1^{-1}\mathbf{C}_{12}\right)^{-1}\mathbf{C}_{12}^T\mathbf{C}_1^{-1} & \left(\mathbf{C}_2-\mathbf{C}_{12}^T\mathbf{C}_1^{-1}\mathbf{C}_{12}\right)^{-1} \\
                                   \end{array}
                                 \right)
\end{eqnarray}
}
confirms $\mathbf{W}=\left(
            \begin{array}{cc}
              \mathbf{W}_{1} & \mathbf{W}_{2} \\
            \end{array}
          \right)$ which completes the proof.$\qquad\blacksquare$
\medskip

\textbf{Proof for universal kriging:} We first compute the kriging weights for the second stage $\mathbf{W}_2$ directly and again use the identity (\ref{blockinv}):
{\footnotesize
\begin{eqnarray*}
 \nonumber % Remove numbering (before each equation)
  \mathbf{W}_2 &=& \left[\left(\mathbf{F}_0-\left(
                                        \begin{array}{c}
                                          \mathbf{C}_{10} \\
                                          \mathbf{C}_{20} \\
                                        \end{array}
                                      \right)^T\left(
                      \begin{array}{cc}
                        \mathbf{C}_1 & \mathbf{C}_{12} \\
                        \mathbf{C}_{12}^T & \mathbf{C}_2 \\
                      \end{array}
                    \right)^{-1}\left(
                                  \begin{array}{c}
                                    \mathbf{F}_1 \\
                                    \mathbf{F}_2 \\
                                  \end{array}
                                \right)\right)\left(\left(
                                        \begin{array}{c}
                                          \mathbf{F}_{1} \\
                                          \mathbf{F}_{2} \\
                                        \end{array}
                                      \right)^T\left(
                      \begin{array}{cc}
                        \mathbf{C}_1 & \mathbf{C}_{12} \\
                        \mathbf{C}_{12}^T & \mathbf{C}_2 \\
                      \end{array}
                    \right)^{-1}\left(
                                  \begin{array}{c}
                                    \mathbf{F}_1 \\
                                    \mathbf{F}_2 \\
                                  \end{array}
                                \right)\right)^{-1}\cdot\right. \\
   & & \left.\cdot\left(
                                        \begin{array}{c}
                                          \mathbf{F}_{1} \\
                                          \mathbf{F}_{2} \\
                                        \end{array}
                                      \right)^T+\left(
                                        \begin{array}{c}
                                          \mathbf{C}_{10} \\
                                          \mathbf{C}_{20} \\
                                        \end{array}
                                      \right)^T\right]\left(
                                                        \begin{array}{c}
                                                          -\mathbf{C}_1^{-1}\mathbf{C}_{12}\left(\mathbf{C}_2-\mathbf{C}_{12}^T\mathbf{C}_1^{-1}\mathbf{C}_{12}\right)^{-1} \\
                                                          \left(\mathbf{C}_2-\mathbf{C}_{12}^T\mathbf{C}_1^{-1}\mathbf{C}_{12}\right)^{-1} \\
                                                        \end{array}
                                                      \right)
\end{eqnarray*}
}
after some matrix manipulations we get
{\footnotesize
\begin{equation}
\begin{array}{r c l}\label{W2direct}
  \mathbf{W}_2 &=& \left(\mathbf{C}_{20}^T-\mathbf{C}_{10}^T\mathbf{C}_1^{-1}\mathbf{C}_{12}\right) \left(\mathbf{C}_{2}-\mathbf{C}_{12}^T\mathbf{C}_1^{-1}\mathbf{C}_{12}\right)^{-1} + \\
  & & +\left(\left(\mathbf{F}_0-\mathbf{C}_{10}^T\mathbf{C}_1^{-1}\mathbf{F}_1\right) - \left(\mathbf{C}_{20}^T-\mathbf{C}_{10}^T\mathbf{C}_1^{-1}\mathbf{C}_{12}\right) \left(\mathbf{C}_{2}-\mathbf{C}_{12}^T\mathbf{C}_1^{-1}\mathbf{C}_{12}\right)^{-1} \left(\mathbf{F}_2-\mathbf{C}_{12}^T\mathbf{C}_1^{-1}\mathbf{F}_1\right)\right)\cdot \\
  & & \cdot\left(\mathbf{F}_1^T\mathbf{C}_1^{-1}\mathbf{F}_1+\left(\mathbf{F}_{2}^T-\mathbf{F}_{1}^T\mathbf{C}_1^{-1}\mathbf{C}_{12}\right) \left(\mathbf{C}_{2}-\mathbf{C}_{12}^T\mathbf{C}_1^{-1}\mathbf{C}_{12}\right)^{-1} \left(\mathbf{F}_2-\mathbf{C}_{12}^T\mathbf{C}_1^{-1}\mathbf{F}_1\right)\right)^{-1}\cdot \\
  & & \cdot\left(\mathbf{F}_{2}^T-\mathbf{F}_{1}^T\mathbf{C}_1^{-1}\mathbf{C}_{12}\right) \left(\mathbf{C}_{2}-\mathbf{C}_{12}^T\mathbf{C}_1^{-1}\mathbf{C}_{12}\right)^{-1}
\end{array}
\end{equation}
}

Now we compute kriging weights and the kriging covariance matrix for the first stage:
Using the notation of (\ref{partitionedmod}) the kriging weights of the first stage are
\begin{eqnarray*}
 \nonumber % Remove numbering (before each equation)
  \mathbf{W}_{12} &=& \left(\left(\mathbf{F}_2-\mathbf{C}_{12}^T\mathbf{C}_1^{-1}\mathbf{F}_1\right)\left(\mathbf{F}_1^T\mathbf{C}_1^{-1}\mathbf{F}_1\right)^{-1}\mathbf{F}_1^T+ \mathbf{C}_{12}^T\right)\mathbf{C}_1^{-1} \\
  \mathbf{W}_{10} &=& \left(\left(\mathbf{F}_0-\mathbf{C}_{10}^T\mathbf{C}_1^{-1}\mathbf{F}_1\right)\left(\mathbf{F}_1^T\mathbf{C}_1^{-1}\mathbf{F}_1\right)^{-1}\mathbf{F}_1^T+ \mathbf{C}_{10}^T\right)\mathbf{C}_1^{-1}
\end{eqnarray*}
which gives the blocks of the kriging covariance matrix of the first stage (\ref{krigcov1})
\begin{eqnarray*}
 \nonumber % Remove numbering (before each equation)
  \boldsymbol\Sigma_2 &=& \left(\mathbf{F}_2-\mathbf{C}_{12}^T\mathbf{C}_1^{-1}\mathbf{F}_1\right)\left(\mathbf{F}_1^T\mathbf{C}_1^{-1}\mathbf{F}_1\right)^{-1}\left(\mathbf{F}_2^T- \mathbf{F}_1^T\mathbf{C}_1^{-1}\mathbf{C}_{12}\right)+\mathbf{C}_2-\mathbf{C}_{12}^T\mathbf{C}_1^{-1}\mathbf{C}_{12} \\
  \boldsymbol\Sigma_{20}^T &=& \left(\mathbf{F}_0-\mathbf{C}_{10}^T\mathbf{C}_1^{-1}\mathbf{F}_1\right)\left(\mathbf{F}_1^T\mathbf{C}_1^{-1}\mathbf{F}_1\right)^{-1} \left(\mathbf{F}_2^T-\mathbf{F}_1^T\mathbf{C}_1^{-1}\mathbf{C}_{12}\right)+\mathbf{C}_{20}-\mathbf{C}_{10}^T\mathbf{C}_1^{-1}\mathbf{C}_{12}
\end{eqnarray*}
Now we apply the update formulae (\ref{updateW22}) to get
\begin{equation}
\begin{array}{r c l}\label{W2update}
  \mathbf{W}_2 &=& \left(\mathbf{C}_{20}-\mathbf{C}_{10}^T\mathbf{C}_1^{-1}\mathbf{C}_{12}+ \left(\mathbf{F}_0-\mathbf{C}_{10}^T\mathbf{C}_1^{-1}\mathbf{F}_1\right)\left(\mathbf{F}_1^T\mathbf{C}_1^{-1}\mathbf{F}_1\right)^{-1} \left(\mathbf{F}_2^T-\mathbf{F}_1^T\mathbf{C}_1^{-1}\mathbf{C}_{12}\right)\right)\cdot \\
   & & \cdot\left(\mathbf{C}_2-\mathbf{C}_{12}^T\mathbf{C}_1^{-1}\mathbf{C}_{12}+ \left(\mathbf{F}_2-\mathbf{C}_{12}^T\mathbf{C}_1^{-1}\mathbf{F}_1\right)\left(\mathbf{F}_1^T\mathbf{C}_1^{-1}\mathbf{F}_1\right)^{-1} \left(\mathbf{F}_2^T-\mathbf{F}_1^T\mathbf{C}_1^{-1}\mathbf{C}_{12}\right)\right)^{-1}
\end{array}
\end{equation}

The inverse in the above equation
\begin{equation}\label{inv1}
  \left(\mathbf{C}_2-\mathbf{C}_{12}^T\mathbf{C}_1^{-1}\mathbf{C}_{12}+ \left(\mathbf{F}_2-\mathbf{C}_{12}^T\mathbf{C}_1^{-1}\mathbf{F}_1\right)\left(\mathbf{F}_1^T\mathbf{C}_1^{-1}\mathbf{F}_1\right)^{-1} \left(\mathbf{F}_2^T-\mathbf{F}_1^T\mathbf{C}_1^{-1}\mathbf{C}_{12}\right)\right)^{-1}
\end{equation}
is the the lower right column block of
\[\left(
    \begin{array}{cc}
      -\mathbf{F}_1^T\mathbf{C}_1^{-1}\mathbf{F}_1 & \mathbf{F}_{2}^T-\mathbf{F}_{1}^T\mathbf{C}_1^{-1}\mathbf{C}_{12} \\
      \mathbf{F}_2-\mathbf{C}_{12}^T\mathbf{C}_1^{-1}\mathbf{F}_1 & \mathbf{C}_{2}-\mathbf{C}_{12}^T\mathbf{C}_1^{-1}\mathbf{C}_{12} \\
    \end{array}
  \right)^{-1}
\]
which may also be computed as
\begin{equation}
\begin{array}{r c l}\label{inv2}
  & & \left(\mathbf{C}_{2}-\mathbf{C}_{12}^T\mathbf{C}_1^{-1}\mathbf{C}_{12}\right)^{-1}- \left(\mathbf{C}_{2}-\mathbf{C}_{12}^T\mathbf{C}_1^{-1}\mathbf{C}_{12}\right)^{-1}\left(\mathbf{F}_2-\mathbf{C}_{12}^T\mathbf{C}_1^{-1}\mathbf{F}_1\right) \cdot \\
  & & \cdot\left(\mathbf{F}_1^T\mathbf{C}_1^{-1}\mathbf{F}_1+\left(\mathbf{F}_{2}^T-\mathbf{F}_{1}^T\mathbf{C}_1^{-1}\mathbf{C}_{12}\right) \left(\mathbf{C}_{2}-\mathbf{C}_{12}^T\mathbf{C}_1^{-1}\mathbf{C}_{12}\right)^{-1} \left(\mathbf{F}_2-\mathbf{C}_{12}^T\mathbf{C}_1^{-1}\mathbf{F}_1\right)\right)^{-1} \cdot \\
  & & \cdot\left(\mathbf{F}_2^T-\mathbf{F}_1^T\mathbf{C}_1^{-1}\mathbf{C}_{12}\right)\left(\mathbf{C}_{2}-\mathbf{C}_{12}^T\mathbf{C}_1^{-1}\mathbf{C}_{12}\right)^{-1}
\end{array}
\end{equation}
In (\ref{W2update}) we now substitute (\ref{inv2}) for (\ref{inv1}) to get
{\footnotesize
\begin{equation*}
\begin{array}{r l}
  \mathbf{W}_2 =& \left(\mathbf{C}_{20}^T-\mathbf{C}_{10}^T\mathbf{C}_1^{-1}\mathbf{C}_{12}\right) \left(\mathbf{C}_{2}-\mathbf{C}_{12}^T\mathbf{C}_1^{-1}\mathbf{C}_{12}\right)^{-1} + \\
  & +\left(\mathbf{F}_0-\mathbf{C}_{10}^T\mathbf{C}_1^{-1}\mathbf{F}_1\right)\left(\mathbf{F}_1^T\mathbf{C}_1^{-1}\mathbf{F}_1\right)^{-1}\textcolor{red}{\clubsuit} \left(\mathbf{F}_{2}^T-\mathbf{F}_{1}^T\mathbf{C}_1^{-1}\mathbf{C}_{12}\right) \left(\mathbf{C}_{2}-\mathbf{C}_{12}^T\mathbf{C}_1^{-1}\mathbf{C}_{12}\right)^{-1}- \\
  & -\left(\mathbf{F}_0-\mathbf{C}_{10}^T\mathbf{C}_1^{-1}\mathbf{F}_1\right)\left(\mathbf{F}_1^T\mathbf{C}_1^{-1}\mathbf{F}_1\right)^{-1} \left(\mathbf{F}_{2}^T-\mathbf{F}_{1}^T\mathbf{C}_1^{-1}\mathbf{C}_{12}\right) \left(\mathbf{C}_{2}-\mathbf{C}_{12}^T\mathbf{C}_1^{-1}\mathbf{C}_{12}\right)^{-1}\left(\mathbf{F}_2-\mathbf{C}_{12}^T\mathbf{C}_1^{-1}\mathbf{F}_1\right)\cdot \\
  & \cdot\left(\mathbf{F}_1^T\mathbf{C}_1^{-1}\mathbf{F}_1+\left(\mathbf{F}_{2}^T-\mathbf{F}_{1}^T\mathbf{C}_1^{-1}\mathbf{C}_{12}\right) \left(\mathbf{C}_{2}-\mathbf{C}_{12}^T\mathbf{C}_1^{-1}\mathbf{C}_{12}\right)^{-1} \left(\mathbf{F}_2-\mathbf{C}_{12}^T\mathbf{C}_1^{-1}\mathbf{F}_1\right)\right)^{-1}\cdot \\
  & \cdot\left(\mathbf{F}_{2}^T-\mathbf{F}_{1}^T\mathbf{C}_1^{-1}\mathbf{C}_{12}\right) \left(\mathbf{C}_{2}-\mathbf{C}_{12}^T\mathbf{C}_1^{-1}\mathbf{C}_{12}\right)^{-1}- \\
  & - \left(\mathbf{C}_{20}^T-\mathbf{C}_{10}^T\mathbf{C}_1^{-1}\mathbf{C}_{12}\right) \left(\mathbf{C}_{2}-\mathbf{C}_{12}^T\mathbf{C}_1^{-1}\mathbf{C}_{12}\right)^{-1} \left(\mathbf{F}_2-\mathbf{C}_{12}^T\mathbf{C}_1^{-1}\mathbf{F}_1\right)\cdot \\
  & \cdot\left(\mathbf{F}_1^T\mathbf{C}_1^{-1}\mathbf{F}_1+\left(\mathbf{F}_{2}^T-\mathbf{F}_{1}^T\mathbf{C}_1^{-1}\mathbf{C}_{12}\right) \left(\mathbf{C}_{2}-\mathbf{C}_{12}^T\mathbf{C}_1^{-1}\mathbf{C}_{12}\right)^{-1} \left(\mathbf{F}_2-\mathbf{C}_{12}^T\mathbf{C}_1^{-1}\mathbf{F}_1\right)\right)^{-1}\cdot \\
  & \cdot\left(\mathbf{F}_{2}^T-\mathbf{F}_{1}^T\mathbf{C}_1^{-1}\mathbf{C}_{12}\right) \left(\mathbf{C}_{2}-\mathbf{C}_{12}^T\mathbf{C}_1^{-1}\mathbf{C}_{12}\right)^{-1}
\end{array}
\end{equation*}
}
At the position $\textcolor{red}{\clubsuit}$ we now multiply with the identity
\begin{equation}
\begin{array}{r c l}
 \mathbf{I} &=& \left(\mathbf{F}_1^T\mathbf{C}_1^{-1}\mathbf{F}_1+\left(\mathbf{F}_{2}^T-\mathbf{F}_{1}^T\mathbf{C}_1^{-1}\mathbf{C}_{12}\right) \left(\mathbf{C}_{2}-\mathbf{C}_{12}^T\mathbf{C}_1^{-1}\mathbf{C}_{12}\right)^{-1} \left(\mathbf{F}_2-\mathbf{C}_{12}^T\mathbf{C}_1^{-1}\mathbf{F}_1\right)\right) \cdot \\
  & & \cdot\left(\mathbf{F}_1^T\mathbf{C}_1^{-1}\mathbf{F}_1+\left(\mathbf{F}_{2}^T-\mathbf{F}_{1}^T\mathbf{C}_1^{-1}\mathbf{C}_{12}\right) \left(\mathbf{C}_{2}-\mathbf{C}_{12}^T\mathbf{C}_1^{-1}\mathbf{C}_{12}\right)^{-1} \left(\mathbf{F}_2-\mathbf{C}_{12}^T\mathbf{C}_1^{-1}\mathbf{F}_1\right)\right)^{-1}
\end{array}
\end{equation}
to get the same $W_2$ as with direct computation (\ref{W2direct}). $\qquad\blacksquare$

\subsection{Proof of the update formula for the kriging covariance matrix of the second stage $\boldsymbol\Sigma_0^+$}
\textbf{Proof:} We compute the kriging covariance matrix for the second stage and plug in the update formulae for the kriging weights (\ref{updateW22}) to get the update formula for the kriging covariance matrix (\ref{updatekrigcov}). The proof is given for the most general setup of universal kriging.

The kriging covariance matrix for the second stage is:
\begin{eqnarray*}
\nonumber % Remove numbering (before each equation)
  \boldsymbol\Sigma_0^+ &=& \left(
            \begin{array}{cc}
              \mathbf{W}_{1} & \mathbf{W}_{2} \\
            \end{array}
          \right)\left(
                   \begin{array}{cc}
                     \mathbf{C}_1 & \mathbf{C}_{12} \\
                     \mathbf{C}_{12}^T & \mathbf{C}_2 \\
                   \end{array}
                 \right)
          \left(
                   \begin{array}{c}
                     \mathbf{W}_{1}^T \\
                     \mathbf{W}_{2}^T \\
                   \end{array}
                 \right)-\left(
            \begin{array}{cc}
              \mathbf{W}_{1} & \mathbf{W}_{2} \\
            \end{array}
          \right)\left(
                   \begin{array}{c}
                     \mathbf{C}_{10} \\
                     \mathbf{C}_{20} \\
                   \end{array}
                 \right)- \\
   & & -\left(
            \begin{array}{cc}
              \mathbf{C}_{10}^T & \mathbf{C}_{20}^T \\
            \end{array}
          \right)\left(
                   \begin{array}{c}
                     \mathbf{W}_{1}^T \\
                     \mathbf{W}_{2}^T \\
                   \end{array}
                 \right)+\mathbf{C}_0= \\
   &=& \mathbf{W}_{1}\mathbf{C}_1\mathbf{W}_{1}^T+\mathbf{W}_{1}\mathbf{C}_{12}\mathbf{W}_{2}^T+\mathbf{W}_{2}\mathbf{C}_{12}^T\mathbf{W}_{1}^T+ \mathbf{W}_{2}\mathbf{C}_2\mathbf{W}_{2}^T- \\
   & & -\mathbf{W}_{1}\mathbf{C}_{10}-\mathbf{W}_{2}\mathbf{C}_{20}-\mathbf{C}_{10}^T\mathbf{W}_{1}^T- \mathbf{C}_{20}^T\mathbf{W}_{2}^T+\mathbf{C}_0
\end{eqnarray*}
Now we plug in the update formula for the kriging weight (\ref{updateW21}) to get:
\begin{eqnarray*}
\nonumber % Remove numbering (before each equation)
  \boldsymbol\Sigma_0^+ &=& \mathbf{W}_{10}\mathbf{C}_1\mathbf{W}_{10}^T-\mathbf{W}_{10}\mathbf{C}_{10}-\mathbf{C}_{10}^T\mathbf{W}_{10}^T+\mathbf{C}_0- \\
   & & -\mathbf{W}_{2}\left(\mathbf{W}_{12}\mathbf{C}_1\mathbf{W}_{10}^T-\mathbf{W}_{12}\mathbf{C}_{10}-\mathbf{C}_{12}^T\mathbf{W}_{10}^T+\mathbf{C}_{20}\right)+ \\
   & & +\mathbf{W}_{2}\left(\mathbf{W}_{12}\mathbf{C}_1\mathbf{W}_{12}^T-\mathbf{W}_{12}\mathbf{C}_{12}-\mathbf{C}_{12}^T\mathbf{W}_{12}^T+\mathbf{C}_{2}\right)\mathbf{W}_{2}^T- \\
   & & -\left(\mathbf{W}_{10}\mathbf{C}_1\mathbf{W}_{12}^T-\mathbf{W}_{10}\mathbf{C}_{12}-\mathbf{C}_{10}^T\mathbf{W}_{12}^T+\mathbf{C}_{20}^T\right)\mathbf{W}_{2}^T
\end{eqnarray*}
Using the kriging covariance matrix for the first stage in the universal kriging setup as in (\ref{krigcovstage1})
{\footnotesize
\begin{eqnarray*}
   & & \left(
    \begin{array}{cc}
      \boldsymbol\Sigma_2 & \boldsymbol\Sigma_{20} \\
      \boldsymbol\Sigma_{20}^T & \boldsymbol\Sigma_0 \\
    \end{array}
  \right)= \\
   &=& \left(
    \begin{array}{cc}
      \mathbf{W}_{12}\mathbf{C}_1\mathbf{W}_{12}^T-\mathbf{W}_{12}\mathbf{C}_{12}-\mathbf{C}_{12}^T\mathbf{W}_{12}^T+\mathbf{C}_{2} & \mathbf{W}_{12}\mathbf{C}_1\mathbf{W}_{10}^T-\mathbf{W}_{12}\mathbf{C}_{10}-\mathbf{C}_{12}^T\mathbf{W}_{10}^T+\mathbf{C}_{20} \\
      \mathbf{W}_{10}\mathbf{C}_1\mathbf{W}_{12}^T-\mathbf{W}_{10}\mathbf{C}_{12}-\mathbf{C}_{10}^T\mathbf{W}_{12}^T+\mathbf{C}_{20}^T & \mathbf{W}_{10}\mathbf{C}_1\mathbf{W}_{10}^T-\mathbf{W}_{10}\mathbf{C}_{10}-\mathbf{C}_{10}^T\mathbf{W}_{10}^T+\mathbf{C}_0 \\
    \end{array}
  \right)
\end{eqnarray*}
}
and $\mathbf{W}_{2} = \boldsymbol\Sigma_{20}^T\boldsymbol\Sigma_2^{-1}$ from (\ref{updateW22}) we get
\begin{eqnarray*}
\nonumber % Remove numbering (before each equation)
  \boldsymbol\Sigma_0^+ &=& \boldsymbol\Sigma_0-\boldsymbol\Sigma_{20}^T\boldsymbol\Sigma_2^{-1}\boldsymbol\Sigma_{20}+ \boldsymbol\Sigma_{20}^T\boldsymbol\Sigma_2^{-1}\boldsymbol\Sigma_2\boldsymbol\Sigma_2^{-1}\boldsymbol\Sigma_{20}- \boldsymbol\Sigma_{20}^T\boldsymbol\Sigma_2^{-1}\boldsymbol\Sigma_{20}= \\
   &=& \boldsymbol\Sigma_0-\boldsymbol\Sigma_{20}^T\boldsymbol\Sigma_2^{-1}\boldsymbol\Sigma_{20}
\end{eqnarray*}
which completes the proof.$\qquad\blacksquare$

%_________________________________________________________________________________________________________________

  \bibliographystyle{elsarticle-harv}
  \bibliography{Bibliography-GVcriterion}

\end{document}